\pdfoutput=1 

\documentclass[10pt,aps,prb,amsmath,amssymb,twocolumn,letterpaper,nobalancelastpage,final,citeautoscript,floatfix,raggedbottom]{revtex4-1}

\usepackage[usenames,dvipsnames]{color}
\usepackage{graphicx}
\usepackage{microtype}
\usepackage{sidecap}
\usepackage[bookmarks=false,colorlinks]{hyperref}
\usepackage{xfrac}
\hypersetup{
    linkcolor=RubineRed,        
    citecolor=ForestGreen,      
    filecolor=Mulberry,     	
    urlcolor=RoyalBlue          
}
\usepackage{natbib}

\newcommand{\gvs}{GaV$_4$S$_8$}
\newcommand{\gvse}{GaV$_4$Se$_8$}
\newcommand{\gms}{GaMo$_4$S$_8$}
\newcommand{\gmse}{GaMo$_4$Se$_8$}
\newcommand{\gns}{GaNb$_4$S$_8$}
\newcommand{\gnse}{GaNb$_4$Se$_8$}
\newcommand{\gts}{GaTa$_4$S$_8$}
\newcommand{\gtse}{GaTa$_4$Se$_8$}

\newcommand{\rw}[1]{\textcolor{black}{#1}}
\newcommand{\revsecond}[1]{\textcolor{black}{#1}}

\begin{document}

\title{Assessing exchange-correlation functional performance in the 
chalcogenide \\ lacunar spinels GaM$_4$Q$_8$ (M = Mo, V, Nb, Ta; Q = S, Se)} 

\author{Yiqun Wang}
  \affiliation{Department of Materials Science and Engineering, Northwestern University, Evanston, Illinois  60208, USA}
  
\author{Danilo Puggioni}
  \affiliation{Department of Materials Science and Engineering, Northwestern University, Evanston, Illinois  60208, USA}
  
\author{James M.\ Rondinelli}
  \email{jrondinelli@northwestern.edu}
  \affiliation{Department of Materials Science and Engineering, Northwestern University, Evanston, Illinois  60208, USA}
  

\begin{abstract}

We perform systematic density functional theory (DFT) calculations to assess the performance of various exchange-correlation potentials $V_{xc}$ in describing the chalcogenide GaM$_4$Q$_8$ lacunar spinels (M=Mo, V, Nb, Ta; Q=S, Se).
We examine the dependency of crystal structure (in cubic and rhombohedral symmetries), electronic structure, magnetism, optical conductivity, 
and lattice dynamics in lacunar spinels at four different levels of $V_{xc}$: the local density approximation (LDA), generalized gradient approximation (GGA), meta-GGA, and hybrid with fractional Fock exchange.
%
\rw{We find that LDA underperforms the Perdew-Burke-Ernzerhof (PBE) and PBE revised for solids (PBEsol) GGA functionals
in predicting lattice constants as well as reasonable electronic structures.}
\rw{The performance of LDA and GGAs can be improved both quantitatively and qualitatively 
by including an on-site Coulomb interaction (LDA/GGA$+U$) 
with a Hubbard $U$ value ranging from 2\,eV to 3\,eV.}
\rw{We find that the PBE functional is able to produce a semiconducting state
in the distorted polar $R3m$ phase 
without on-site Coulomb interactions.}
%
%
The meta-GGA functional SCAN predicts reasonable lattice constants and electronic structures; it exhibits behavior similar to the GGA$+U$ functionals for small $U$ values \rw{of 1\,eV to 2\,eV}. 
The hybrid functional HSE06 is accurate in predicting the lattice constants, 
\rw{but leads to a band gap
greater than the experimental estimation of 0.2\,eV\cite{corraze2013electric,cario2010electric} in this family.}
All of the lacunar spinels in the cubic phase are metallic at
these levels of band theory, however, the predicted valence bandwidths are extremely narrow ($\approx$0.5\,eV).
\rw{The DFT ground states of cubic vanadium chalcogenides are found to be highly spin-polarized, which contrast previous experimental results.}
\rw{With spin-orbit coupling (SOC) interactions and a Hubbard $U$ value of 2\,eV to 3\,eV, we predict a semiconducting cubic phase in all compounds studied.}
\rw{SOC does not strongly impact the electronic structures of the symmetry-broken $R3m$ phase.}
We also find that these $V_{xc}$ potentials do not quantitatively agree with the available experimental optical conductivity on  \gvs{}; nonetheless, the LDA and GGA functionals correctly reproduce its lattice dynamical modes.
Our findings suggest that accurate qualitative and quantitative 
simulations of the lacunar spinel family with DFT requires careful 
attention to the nuances of the exchange-correlation functional and considered spin structures.
\end{abstract}

\maketitle


\section{Introduction}

The lacunar spinel family GaM$_4$Q$_8$ (M = Mo, V, Nb, Ta; Q = S,Se) 
have garnered attention for decades owing to their fascinating 
properties, which include 
metal-insulator transitions\cite{camjayi2014first}, 
the capability to host skyrmion lattices\cite{kezsmarki2015neel}, 
and multiferroism\cite{widmann2017multiferroic}.
\gvs{} and \gms{} are the most well studied materials in this family; 
they exhibit Jahn-Teller-type structural phase transitions at $\approx$40\,K 
upon cooling, followed by spontaneous magnetic ordering below their Curie temperatures 
$T_\mathrm{C}$ \cite{pocha2000electronic}.
The multiple phase transitions -- 
metallic-to-insulating and  paramagnetic-to-ferromagnetic -- 
connecting distinct physical states 
make these transition-metal compounds ideal candidate materials for novel electronic platforms\cite{cario2010electric}.

After decades of continuous studies on various properties of the lacunar spinel compounds, 
the mechanism of these phase transitions as well as the proper theoretical 
approaches to describe various electronic states are still unclear.
For instance, while the vanadium and molybdenum compounds can undergo symmetry-lowering structural phase transitions  
at low temperature\cite{sahoo1993evidence, francois1991structural, franccois1992structural}, the 
niobium and tantalum lacunar spinels remain in the high-symmetry cubic phase over a broad temperature range\cite{abd2004transition}.
One of the possible reasons for this behavior in the family may be attributed 
to variations in the strength of electron-electron interactions,\cite{sahoo1993evidence} 
since electron-correlation effects are expected to be stronger in 
3$d$ rather than 5$d$ transition metals. 
However, there is also evidence that local structural distortions in 
lacunar spinel compounds could lead to insulating states even in the 
absence of strong correlation\cite{sieberer2007importance}.
In addition, a number of members within the lacunar spinel family exhibit interesting resistive-switching behavior \cite{corraze2013electric},
making them potential materials for resistive random-access memory (RRAM) materials.
Much of the literature attributes the aforementioned features to  
the special tetrahedral transition-metal clusters within the unit cell\cite{le1995tetrahedral}; yet, how and why it supports all of these 
properties remains to be agreed upon. \cite{rastogi1984electron,rastogi1987magnetic,pocha2000electronic}
In order to have a better understanding of the structure-property relationships
among the  lacunar spinels, a qualitative and possibly quantitative investigation of
electron-correlation effects and structural distortions within these 
materials is needed.

Density functional theory (DFT) simulations are widely used in solid-state materials research 
owing to the efficiency and accuracy they achieve by replacing the original many-electron interaction problem with an auxiliary independent-particle problem through 
a suitably constructed exchange-correlation potential ($V_{xc}$).
Because DFT simulations can capture the interplay of structural effects on electron-electron interactions and its dependence on determining the ground state, it is an ideal method to study the lacunar spinels with many 
internal atomic, spin, and orbital degrees-of-freedoms.
However, no available $V_{xc}$ can provide the exact description of exchange and correlation, which necessitates benchmarking both common and state-of-the-art 
density functionals against available experimental data.
To that end, it becomes possible to identify 
the optimal functional for  describing and predicting properties in 
the lacunar spinel family.

In this work, we systematically benchmark the performance of DFT $V_{xc}$ functionals 
in describing the lacunar spinel family at four rungs of ``Jacob's ladder'', 
specifically the local density approximation (LDA), the 
generalized gradient approximation (GGA) as implemented by Perdew-Burke-Ernzerhof (PBE) 
and PBE revised for solids (PBEsol), the meta-GGA functional SCAN, and 
the hybrid functional HSE06.
Our aim is to identify 
the best description of the lacunar spinel family 
from first-principles DFT simulations and where compromises on performance 
are made so as to to facilitate future studies and predictions.
To that end, we investigate the functional dependency of lattice parameters, magnetism, electronic structures, optical properties, 
and lattice dynamics in both the cubic and Jahn-Teller 
distorted rhombohedral phases.
Our main conclusion is that GGA and higher level $V_{xc}$ functionals 
\rw{are more reasonable than} 
LDA in predicting almost all properties assessed. 
The GGA functionals with an on-site Coulomb interaction (GGA$+U$) 
values of $U\approx2$\,eV quantitatively improves functional performance for 
the electronic structures of the rhombohedral phases.
\rw{Spin-orbit coupling (SOC) interactions lift orbital degeneracies in the electronic structures of the cubic phases and enable a semiconducting phase to emerge with Hubbard $U$ values ranging from 2\,eV to 3\,eV. However, SOC does not significantly impact the electronic structures of the rhombohedral phase, where orbital symmetry is already broken by lattice distortions.}
SCAN and HSE06 are able to predict accurate lattice parameters, but HSE06 leads to band gaps significantly larger than experimental estimations.
Our findings suggest that the predicted physical properties of the lacunar spinel family are highly $V_{xc}$ functional dependent.
Therefore, it is important to benchmark different $V_{xc}$ performance on properties of interest before further studies.
It is likely that the coupling of internal degrees of freedom in lacunar spinels, 
e.g., local cluster distortion, intra- and inter-cluster electronic and magnetic interactions,
underlie the observed fascinating behavior as well as our reported 
high sensitivity to $V_{xc}$ in this materials family.
%


\section{Materials and Methods}

\subsection{Crystal structure, electrical, and magnetic properties}

\begin{figure}
  \center
  \vspace{-1mm}
 \includegraphics[width=0.9\linewidth]{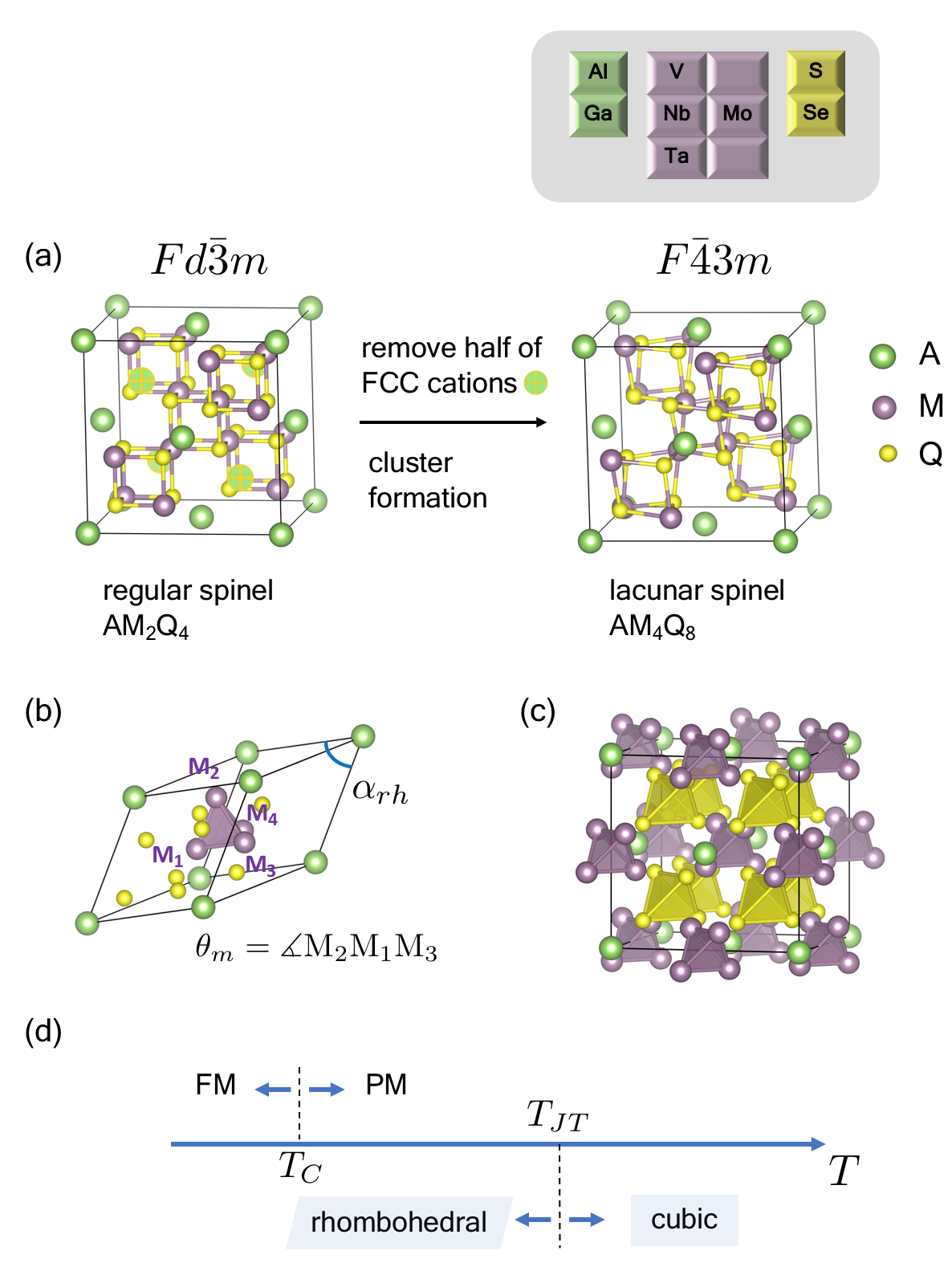}
  \caption{(a) Derivation of the cubic phase lacunar spinel AM$_4$Q$_8$ from an ideal spinel structure, some anions are hidden in the figure to facilitate visualization of the M$_4$ cluster formation. (b) The primitive cell of AM$_4$Q$_8$ in both cubic and rhombohedral phases, with the interaxial rhombohedral angle $\alpha_{rh}$, intra-cluster metal-metal-bond angle $\theta_m$. (c) M$_4$ cluster connectivity in the cubic phase, they occupy the four octahedral holes created by the A cations. (d) Schematic phase diagram of lacunar spinels exhibiting multiple phase transitions. (Key: FM=ferromagnetic, PM=paramagnetic).}
  \label{fig:fig1}
\end{figure}

The crystal structure of the lacunar spinel, also referred to as an 
A-site deficient spinel (AM$_4$Q$_8$), is derived from the regular spinel 
(AM$_2$Q$_4$ composition) by removing the interpenetrating FCC A-site 
sublattice as depicted in \autoref{fig:fig1}(a).
Upon removing half of the A-site cations occupying the tetrahedral holes in 
the regular spinel,
the space group loses inversion symmetry, reducing from $Fd\bar{3}m$ to $F\bar{4}3m$ (space group no.\ 216).
The structure then undergoes additional internal displacements and 
spontaneous strains: 
the previously equidistant M-M network breaks into isolated tetrahedral 
transition-metal clusters with chalcogenide ligands [M$_4$Q$_4$]$^{5+}$.
In order to quantitatively describe the internal degrees of freedom in the crystal structure, 
we define $\alpha_{rh}$ as the interaxial angle of the rhombohedral unit cell, 
and $\theta_{m}$ as the M$_2$-M$_1$-M$_3$ angle centering the apical metal atom 
along the $C_{3v}$ axis of the M$_4$ cluster, as shown in \autoref{fig:fig1}(b).

\begin{table*}
\begin{ruledtabular}
\caption{Experimental Jahn-Teller ($T_\mathrm{JT}$) and Curie ($T_\mathrm{C}$) 
transition temperatures and unit cell volumes (V) for the distorted $R3m$ 
vanadium and molybdenum lacunar spinels. 
The vanadium (molybdenum) chalcogenides  exhibit acute (obtuse) angluar 
distortions away from the ideal cubic 60$^\circ$.
$\theta_m$ and $\alpha_{rh}$ are obtained at temperatures below $T_\mathrm{JT}$,
and no significant structural changes have been observed at around $T_\mathrm{C}$.
\label{tab:lattice_params}}
\centering
\begin{tabular}{llllllll}
Compound & $T_\mathrm{JT}$\,(K) & $T_\mathrm{C}$\,(K) & $\alpha_{rh}$\,($^{\circ}$) &  $\theta_m$\,($^{\circ}$) & V$_{\mbox{cell}}^{F\bar{4}3m}$ (\AA$^{3}$) & V$_{\mbox{cell}}^{R3m}$ (\AA$^{3}$) & Ref.\ \\
\hline
\gvs & 44 & 12.7 & 59.6  & 58.4 & 225.6 & 224.3 & \onlinecite{widmann2017multiferroic},
    \onlinecite{powell2007cation} \\
\gvse & 41 & 17.5  & 59.6 & 57.7 & 260.7 & 259.6 & \onlinecite{bichler2010magnetismus},\onlinecite{fujima2017thermodynamically}  \\
\hline
\gms & 45 & 19.5 & 60.5 & 61.6 & 230.1 & 230.0 & \onlinecite{francois1991structural},\onlinecite{powell2007cation} \\
\gmse & 45 & 23 & 60.6 & 61.4 & 263.3 & 262.2 & \onlinecite{franccois1992structural}
\end{tabular}
\end{ruledtabular}
\end{table*}

At room temperature, all lacunar spinels studied here exhibit 
cubic $F\bar{4}3m$ symmetry with $\alpha_{rh} = \theta_{m} = 60^{\circ}$.
\gvs, \gvse, \gms, and \gmse, however, 
undergo symmetry-lowering structural Jahn-Teller (JT) transitions 
at $\approx$40\,K from $F\bar{4}3m$ to $R3m$ (space group no.\ 160) 
\cite{sahoo1993evidence, francois1991structural, franccois1992structural}, 
followed by spontaneous magnetic ordering at a lower Curie temperature $T_\mathrm{C}$ 
[\autoref{fig:fig1}(d)].
These displacive distortions lead to a unit cell of slightly different volume, lattice parameters, and rhombohedral angle $\alpha_{rh}$. 
The geometry of the metal cluster within the unit cell is also distorted away from its cubic structure and the 
$\theta_{m}$ angle deviates from the ideal cubic value (60$^\circ$).
The experimental crystallographic data for the distorted vanadium and molybdenum lacunar spinels are tabulated in \autoref{tab:lattice_params},
where we also provide the Jahn-Teller and Curie transition temperatures. 
%

\begin{figure}
\center
\includegraphics[width=0.9\linewidth]{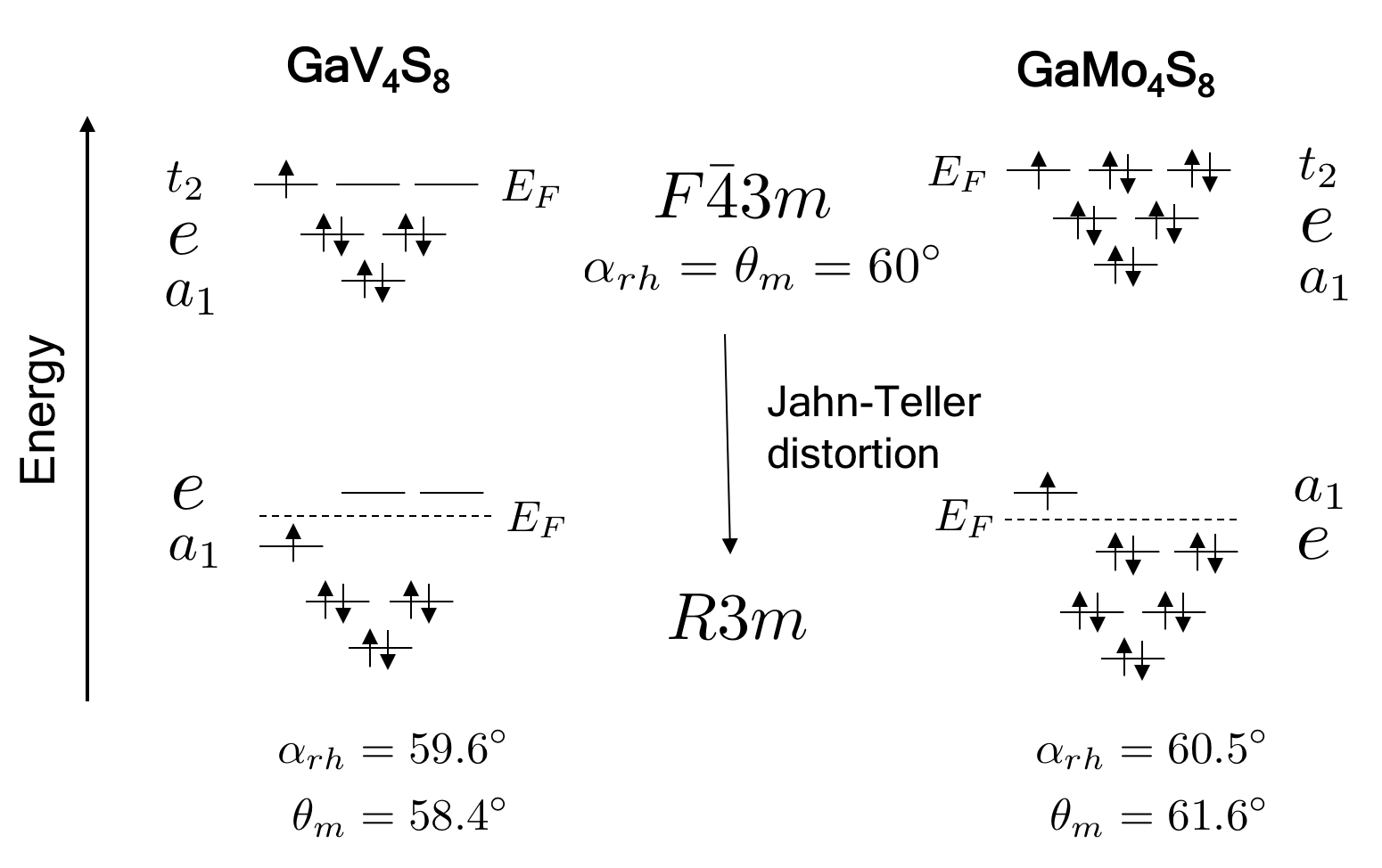}
\caption{Valence molecular orbital diagrams of \gvs{} and \gms. The valence $t_2$-symmetry 
orbitals are triply degenerate in the cubic phase. The orbital degeneracy is 
lifted by the accompanied Jahn-Teller distortion with a distortion 
sense that stabilizes and leads to filling of either the $a_1$ (\gvs) or $e$ (\gms) orbitals based on orbital occupancy of the metals forming the cluster. The dotted lines indicate the Fermi level in the distorted phases, whereas in the cubic phase the Fermi level intersects the triply degenerate valence bands.
}
  \label{fig:fig2}
\end{figure}

The lacunar spinels are reported to be narrow-bandwidth semiconductors 
with $\approx$0.2\,eV band gaps that vary with temperature\cite{pocha2000electronic,cario2010electric,widmann2017multiferroic}.
Early work showed that the valence bands mainly consist of transition-metal $d$ orbitals\cite{shanthi1999electronic}.
Since the transition-metal M$_4$ clusters are relatively distant from each other 
with about 4 \AA{} inter-cluster separation [\autoref{fig:fig1}(c)], the low-energy valence 
electronic structure can be described using a molecular orbital (MO) diagram 
for the cluster (\autoref{fig:fig2}).
In the cubic phase, the valance bands are triply degenerate with $t_{2}$ symmetry. 
V, Nb, and Ta chalcogenide lacunar spinels all exhibit $t_2^1$ occupancies 
whereas Mo exhibits $t_2^5$ filling, indicating susceptibility to a first-order 
Jahn-Teller distortion.
After the Jahn-Teller structural distortion, the triply degenerate $t_2$ orbital splits into two sets of orbitals, 
$a_1$ and $e$.
The relative energy of the two sets of orbitals is occupancy-dependent; the 
$a_1$ orbital is more stable in the vanadium compounds whereas the $e$ orbitals 
are preferentially stabilized in the molybdenum compounds.

In addition to these structural transitions, the vanadium and molybdenum compounds 
show spontaneous magnetic ordering at $T_\mathrm{C}$ when in the rhombohedral phase.
\gvs{} is also reported to have a complex magnetic phase diagram at low temperature\cite{kezsmarki2015neel}.
The effective local magnetic moment in both the paramagnetic and ferromagnetic phases corresponds to approximately 1 unpaired electron per unit cell, and is mostly localized about the transition-metal cluster\cite{pocha2000electronic} rather than on 
the individual atomic sites comprising the cluster. 
\gns, \gnse, \gtse{} are paramagnetic at ambient conditions with effective magnetic moments of 1.76\,$\mu_B$, 1.6\,$\mu_B$, and 0.7\,$\mu_B$ per cluster\cite{pocha2005crystal,abd2004transition}. No structural phase transition or spontaneous magnetic ordering are reported in these compounds down to 1.6\,K\cite{abd2004transition}. 
Last, we note that the family of materials is also often referred to as Mott insulators owing 
to the large distance between transition-metal clusters\cite{pocha2000electronic} and 
not typically because of strong electron-electron interactions\cite{sieberer2007importance} although they 
likely play some role.
The semiconducting behavior is typically attributed to variable range hopping (VRH) conduction\cite{sahoo1993evidence} among these separated metal clusters.
Nonetheless, the microscopic mechanisms behind the semiconducting nature, 
as well as the multiple phase transitions, are still under active investigation\cite{widmann2017multiferroic,reschke2017optical}.


\subsection{Exchange-correlation functionals}
We use exchange-correlation potentials ($V_{xc}$) at four different levels of approximation to assess the structure and properties of the 
chalcognide lacunar spinels. 
The functionals examined include LDA, 
GGA as implemented by Perdew-Burke-Ernzerhof (PBE)\cite{perdew1996jp}, 
and PBE revised for solids (PBEsol)\cite{csonka2009assessing}, 
meta-GGA functional SCAN as implemented by Sun et al.,\cite{sun2015strongly} 
and Heyd-Scuseria-Ernzerhof hybrid functional HSE06\cite{heyd2005energy}.
The $V_{xc}$ in LDA is not derived from first principles, but from Monte Carlo simulations of the uniform electron gas.
The functional solely depends upon the local electron density in space and usually provides a good approximation for simple materials (including metals)
with electronic states that vary slowly in space. 
However, the LDA potentials decay rapidly for finite systems while the 
true exchange-correlation potential has significant non-local contributions; 
this behavior often leads to overestimation of the binding energy\cite{van1999correcting} 
and underestimation of lattice constants in solids\cite{haas2009calculation}.

To improve on the LDA, GGA functionals that take the gradient of electron 
density $\nabla n(\mathbf{r})$ into consideration have been developed. 
The PBE and PBEsol functionals 
improve the binding energy by roughly an order of magnitude, 
but have a general tendency to overestimate lattice constants\cite{haas2009calculation}.
Since LDA and GGA functionals are well-known to be unable to predict the 
insulating state of Mott insulators\cite{paier2006screened} with strong correlations 
and nonlocal exchange, the beyond DFT method, DFT$+U$, is typically used 
to account for such interactions among the localized $d$ electrons. 
The on-site Coulomb interaction term $U$ favors the on-site occupancy matrix 
towards fillings that are fully occupied or fully unoccupied and hence a 
more localized electronic structure within the correlated manifold.
Here, we use the GGA functionals PBE and PBEsol with on-site Coulomb interaction 
(GGA$+U$) with $U$ values of 1.0, 2.0 and 3.0 eV on the the M-metal sites using 
the formalism introduced by Dudarev et al.
\cite{dudarev1998electron} to assess the effect of electron correlation in the M$_4$ clusters.
\rw{The range of $U$ values is based on results from  previous computational studies}\cite{zhang2017magnetic,muller2006magnetic,pocha2005crystal} \rw{and our own preliminary assessments, where reasonable we focused on band gap and magnetic moment predictions.}

It comes naturally from the previous two rungs of Jacob's ladder that the second-order derivative of the electron density should be considered.
Meta-GGA functionals are essentially an extension to GGAs whereby the Laplacian of the electron density $\nabla^{2} n(\mathbf{r})$ is also considered. 
In practice, the kinetic energy density 
$\tau(\mathbf{r}) = \sum_{i=1}^{N_{occ}}\frac{1}{2}\left| \nabla\psi_{i}(\mathbf{r})\right|^{2}$ is used, where the summation runs over the occupied Kohn-Sham orbitals 
$\psi_{i}(\mathbf{r})$.
The recently developed meta-GGA functional SCAN 
(strongly constrained and appropriately normed semi-local density function) 
fulfills all known constraints required by the exact density functional, 
and is reported to have achieved remarkable accuracy for systems where the exact exchange-correlation hole is localized around its electron\cite{sun2015strongly}.

Hybrid DFT functionals incorporate a portion of exact exchange interaction from Hartree-Fock (HF) theory 
with that of a local or semi-local density functional.
The semi-empirical hybrid functional B3LYP has been widely used for 
finite chemical systems and shown more accurate results in thermochemical 
and electronic properties\cite{tirado2008performance,di2006electronic}.
In periodic solid state systems, one route to incorporate 
an exact exchange interaction is by means of range separation.
In the range separated HSE06 hybrid functional, the short-range (SR) exchange interaction consists of partial contributions from exact exchange and the PBE functional. 
The long-range (LR) part of the Fock exchange term is replaced by that from 
the semi-local PBE functional.
The correlation term from PBE is used in the HSE06 hybrid functional.
The resulting exchange-correlation energy expression is 
\[
	E^\mathrm{HSE06}_{xc} = \frac{1}{4}E^\mathrm{HF,SR}_x + \frac{3}{4}E^\mathrm{PBE, SR}_x + E^\mathrm{PBE, LR}_x + E^\mathrm{PBE}_c
\]

The inclusion of exact-exchange interactions in hybrid functionals also partly fixes the self-interaction problem in pure DFT functionals, 
and can provide accurate descriptions of lattice parameters, 
bulk moduli and band gaps in periodic
systems\cite{he2012screened,hummer2009heyd,ramzan2013electronic}.


\subsection{Computational details}
We perform DFT simulations as implemented in the Vienna Ab initio Simulation Package (VASP)\cite{kresse1996efficient, kresse1999ultrasoft}.
The projector augmented-wave (PAW) potentials\cite{blochl1994projector} are used for all elements in our calculations with the following  
valence electron configurations:
Ga ($3d^{10}4s^{2}4p^{1}$), Mo ($4s^24p^64d^55s^1$), V ($3s^23p^63d^44s^1$), Nb ($4s^24p^64d^45s^1$), 
Ta ($5p^65d^46s^1$), S ($3s^23p^4$), and Se ($4s^24p^4$).
Based on convergence test with respect to $k$-point meshes in reciprocal space and plane wave basis set cutoff energies, 
we use a $\Gamma$-centered $6\times6\times 6$ mesh with a 500\,eV kinetic energy cutoff.
For HSE06 calculations, we use a $4\times4\times 4$ $k$-point mesh and a 
400\,eV kinetic energy cutoff 
due to the high computational cost and convergence difficulties for 
the spin-polarized calculations.
Since the lacunar spinels are small-gap semiconductors, we employ 
Gaussian smearing with a small 0.05\,eV width. 
For density-of-state calculations, we use the tetrahedron method 
with Bl\"{o}chl corrections.\cite{PhysRevB.49.16223}

We perform full lattice relaxations with different DFT functionals 
until the residual forces on an individual atom are less than 1.0\,meV\AA$^{-1}$.
The experimental crystal structures of the lacunar spinels \gvs, \gvse, \gms, \gmse, \gns, \gnse, and \gtse{} 
are obtained from the Inorganic Crystal Structure Database (ICSD) \cite{bergerhoff1987crystallographic} and used 
as initial inputs for these geometry relaxations.
\gts{} is not included here since the structure is not experimentally reported. 
%
\rw{Both high-temperature cubic  and low-temperature rhombohedral phases are investigated for all target compounds.}
\rw{Crystal structures of the rhombohedral phase Nb and Ta compounds are obtained by making a small displacement to their cubic atomic positions along the symmetry-lowering pathway (i.e., from $F\bar{4}3m$ to $R3m$), followed by DFT structural relaxations.}
All \rw{experimental and DFT-relaxed} crystal structures are available electronically at Ref.\, \onlinecite{github_link}.
The effect of on-site Coulomb interactions on the crystal structures is
also investigated at the \rw{LDA and} GGA functional level. 
Since the lacunar spinels exhibit various magnetic properties, 
we also initialize the calculations with multiple possible magnetic configurations for the lattice relaxations.
This is a necessary process owing to the multiple metastable spin configurations accessible.
The magnetic configuration with the lowest energy is reported as the DFT ground state and used to compare with other functional results.
Spin-orbit interactions are \rw{also considered in our electronic structure simulations 
owing to their potentially significant impact on the orbital structure of 4$d$ and 5$d$ transition metals.}
\revsecond{For spin-orbit coupling (SOC) calculations, we use the fully-relaxed crystal structures from the aforementioned non-SOC simulations. The magnetic moment is set to be 1$\mu_B$ per formula unit along the (111) direction for both the cubic and rhombohedral phases. }

\rw{Zone center ($\mathbf{k}=\mathbf{0}$)} phonon frequencies 
and eigendisplacements for both the cubic and rhombohedral phases of \gvs{} 
(within primitive cells) 
are obtained using the frozen-phonon method 
with pre- and post-processing performed with the Phonopy package\cite{phonopy}.


\section{Results and discussions\label{sec:results}}


\subsection{$F\bar{4}3m$ cubic phase}


\subsubsection{Lattice parameters}

\begin{figure}
  \centering
 \includegraphics[width=0.99\columnwidth]{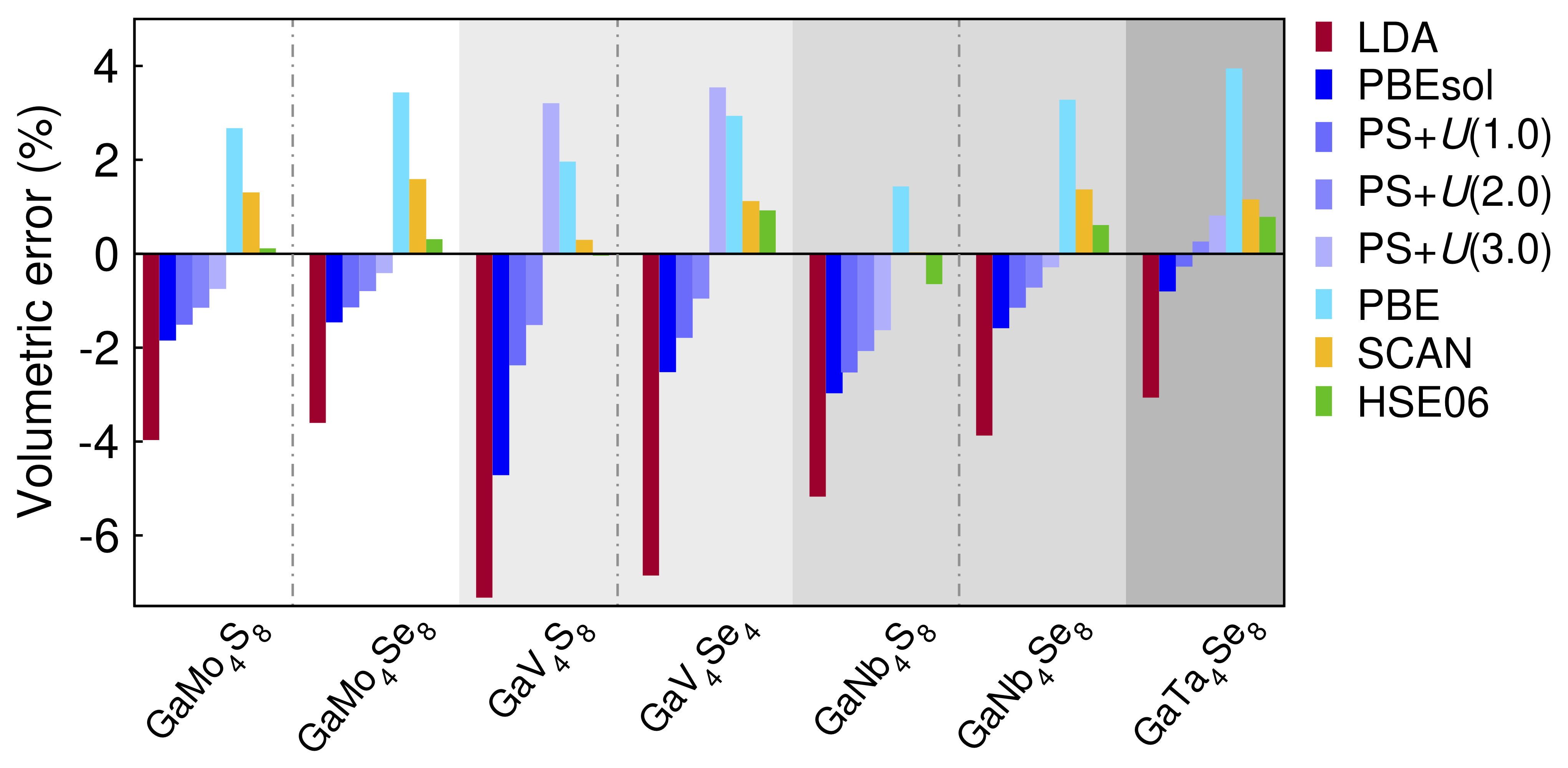}
  \caption{Relative error in the unit cell volume of the cubic phase at different  levels of DFT. PS is an abbreviation for the PBEsol functional and the number in parenthesis is the value of the on-site Coulomb interaction used in the GGA$+U$ method.}
  \label{fig:fig3}
\end{figure}

The crystal structures of the lacunar spinels with cubic symmetry are fully relaxed with DFT using the different $V_{xc}$ potentials.
\autoref{fig:fig3} shows the volumetric error for the DFT ground state unit cell volume relative to the experimental room temperature data. 
For molybdenum and vanadium compounds, we report the cell volumes of 
ferromagnetic spin structures with magnetic moments of $1\,\mu_B$ and $5\,\mu_B$ per formula unit, respectively.
The niobium and tantalum compounds are non-magnetic at all DFT functionals levels.
See \autoref{sec:mag_F43m} for the 
detailed descriptions of the magnetic moment configurations.

In general, the LDA and PBEsol functionals underestimate the lattice parameters, 
while PBE predicts larger lattice constants compared with experimental data. 
LDA has relatively larger deviations (4\,\% or higher) compared with the GGA results,
\rw{it is a well-known problem that LDA tends to underestimate the lattice constants.} 
%
We also check the effect of on-site Coulomb interactions (\rw{LDA and GGA$+U$}) on lattice parameters with $U$ values up to 3\,eV. 
With increasing \rw{on-site Coulomb interaction strength}, we find 
the lattice parameters follow a monotonic increasing trend \rw{for both the LDA and GGA functionals. We only show the trend for PBEsol in \autoref{fig:fig3} owing to its similarity with the others.}
\rw{Therefore, a reasonable Hubbard $U$ value could quantitatively improve the lattice parameter predictions in the LDA and PBEsol functionals.}
%
%

Interestingly, the vanadium compounds exhibit cell volumes that are 
the most sensitive to the choice of the $U$ value among the 
lacunar spinels.
For instance, the difference in volumetric error induced by $U=3.0$\,eV 
is less than 2\,\% in \gms, but is almost 8\,\% in \gvs.
The highly spin-polarized electronic state used for the vanadium compounds may 
be a possible cause of the different sensitivity on the on-site Coulomb interactions.

GGA functionals with 
$U<3.0$\,eV generally predict reasonable cubic lattice constants 
with less than 4\,\% error in the cell volumes.
The meta-GGA functional SCAN and hybrid functional HSE06 have smaller errors in predicting lattice constants, 
which give less than 2\% error for all 7 compounds studied here.
Considering the high computational cost of structural relaxations with HSE06, SCAN should be preferred over HSE06 
for lattice parameter estimation unless one requires a specific accuracy requirement or improved forces.

These results suggest that most of the DFT functionals are able to predict reasonable cubic phase
crystal structures in the lacunar spinel family with less than 4\,\% error in the volumes. 
%
%
%
\rw{Generally, we recommend using GGA functionals with a tunable Hubbard $U$ value of 1\,eV to 3\,eV for lattice parameter predictions.}
SCAN and HSE06 give more accurate lattice constants compared with lower-level functionals, 
while SCAN is preferable based on a compromise between accuracy and efficiency.


\subsubsection{Magnetism}\label{sec:mag_F43m}

Experimentally, the vanadium and molybdenum compounds exhibit paramagnetism 
above their Curie temperatures and exhibit spontaneous magnetic ordering at 
low temperature\cite{pocha2000electronic}. 
The magnetically ordered phases can host multiple fascinating magnetic states, 
including ferromagnetism and complex spin textures (skyrmion lattices) 
\cite{kezsmarki2015neel}. 
Those complex magnetic structures are not considered here. 
The niobium and tantalum compounds show very weak magnetism and do not exhibit spontaneous magnetic ordering down to 1.6K\cite{abd2004transition}.
Since the transition-metal clusters are relatively far from each other with a 
distance of around 4\,\AA, 
the inter-cluster magnetic interactions are expected to be quite small. 
Here we use a ferromagnetic spin configuration on all metal sites within the 
cluster to model the magnetically ordered phases. 

From our DFT simulations, different transition metal clusters are able to hold various magnetic configurations. 
For the molybdenum compounds, we are only able to stabilize one ferromagnetic configuration in the cubic phase 
which corresponds to 1\,$\mu_B$ per primitive cell. The magnetic moments are evenly distributed about the four molybdenum atoms in the Mo$_4$ cluster with negligible contributions from other atomic species. 
In contrast, the vanadium compounds show numerous stable magnetic configurations (\autoref{table:mag_energy_cubic}). 
Apart from the same ferromagnetic configuration as in the molybdenum compounds, we also find a highly spin-polarized state in the cubic phase.
To the best of our knowledge, \rw{the electronic structures of this state has not yet been reported before.}
Recent neutron diffraction studies show that there is one single spin distributed across the V$_4$ cluster instead of residing on a single vanadium ion\cite{JeffLynn}.

In the highly spin-polarized state, the magnetic moment could be $\,5\mu_B$ or  $7\,\mu_B$ per formula unit (f.u.), depending on the DFT functional used.
The spin-moments are evenly distributed about the transition-metal cluster, 
with approximately 1.25\,$\mu_B$ magnetic moment localized on each vanadium atom.
This state is significantly lower in energy than the ferromagnetic configuration with 1\,$\mu_B$ per formula unit in our DFT simulations.

\begin{table}
\begin{ruledtabular}
\centering
\caption{\label{table:mag_energy_cubic} Energy differences (in eV/f.u.) of different magnetic  configurations compared with non-magnetic calculations for the cubic phase.
$E_{\sigma}$ denotes the energy of the highly-polarized state with 5\,$\mu_B$ or 7\,$\mu_B$ per formula unit.
An `--' indicates that the state was not stable. \revsecond{PS is an abbreviation for the PBEsol functional and the number in parenthesis is the value of the on-site Coulomb interaction used in the GGA$+U$ method.}}
\begin{tabular}{lcccc}%
 & \multicolumn{2}{c}{\gvs} & \multicolumn{2}{c}{\gvse} \\
\cline{2-3}\cline{4-5}
 & $E_{\sigma}$	& $E_{\mu=1\mu_B}$	& $E_{\sigma}$ & $E_{\mu=1\mu_B}$ \\
\hline
LDA & -- & -- & -- & -- \\
LDA$+U(1.0)$ & -- & 0.005 & -0.172 & -0.007  \\
LDA$+U(2.0)$ & -0.404 & -0.008 & -0.68 & 0.06 \\
LDA$+U(3.0)$ & -0.946 & -0.03 & -1.264 & -0.051 \\
\hline
PBEsol	& -- & -0.001 & -0.072 & -0.008 \\
\hline
PS$+U(1.0)$ & -0.305 & -0.014 & -0.562 & -0.028 \\
PS$+U(2.0)$ & -0.823 & -0.035 & -1.123 & -- \\
PS$+U(3.0)$ & -1.641 & -- & -2.174 & -- \\
\hline
PBE & -0.095 & -0.010 & -0.327 & -0.022 \\
\hline
SCAN & -0.875 & -0.049 & -1.196 & -0.064 \\
\hline
HSE06 & -1.355 & 0.005 & -1.742 & -0.092 \\
\end{tabular}
\end{ruledtabular}
\end{table}

\begin{figure}
  \centering
 \includegraphics[width=0.95\linewidth]{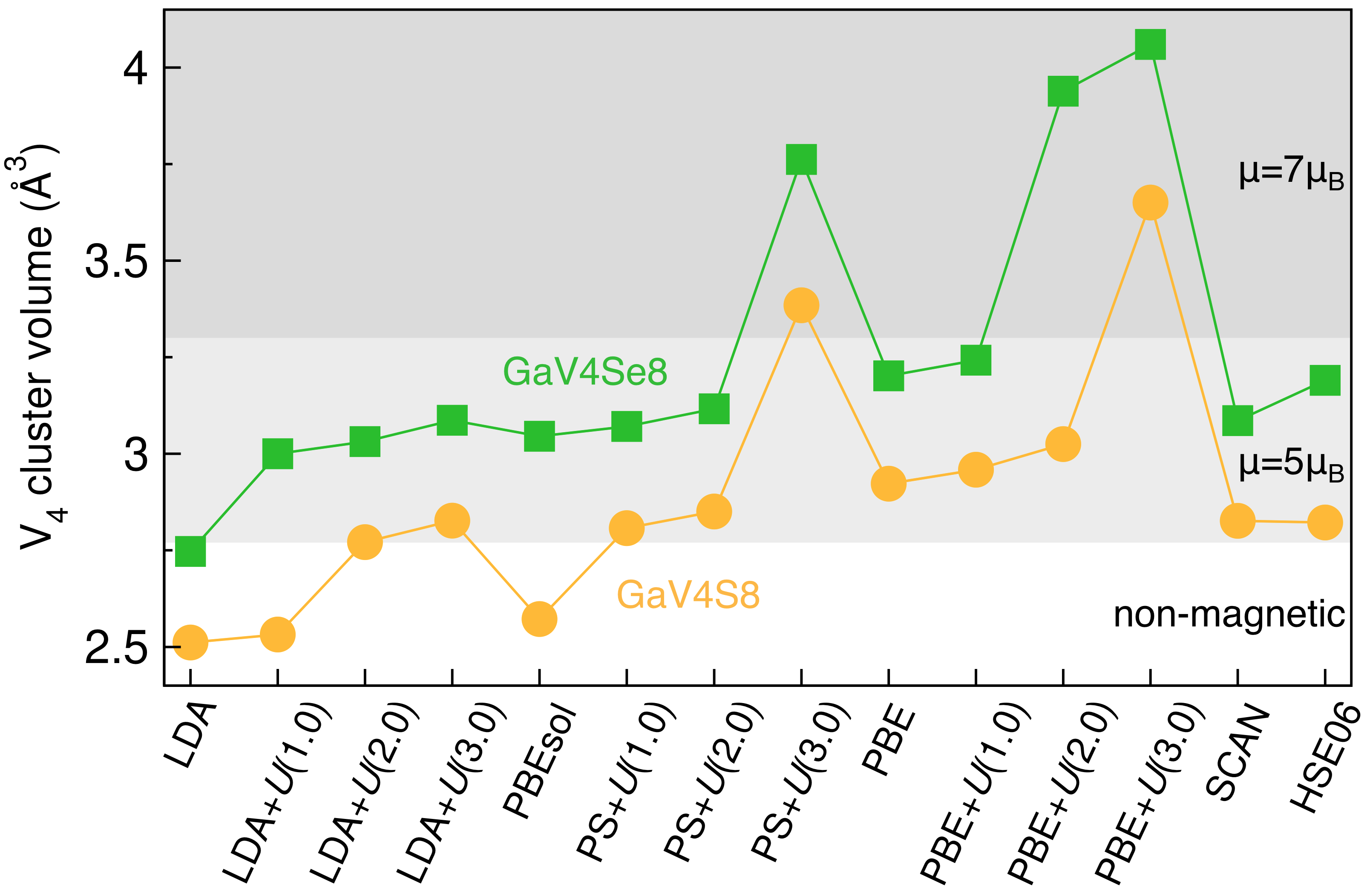}
  \caption{The volume of the tetrahedral V$_4$ cluster with 
  different DFT functionals and their corresponding ground state magnetic moment per formula unit.
  The white area shows non-magnetic results. The light-shaded and dark-shaded areas correspond to states with 5\,$\mu_B$ and 7\,$\mu_B$ magnetic moments, respectively. \revsecond{PS is an abbreviation for the PBEsol functional and the number in parenthesis is the value of the on-site Coulomb interaction used in the GGA$+U$ method.}}
  \label{fig:fig4}
\end{figure}
 
\begin{figure*}[tbh]
  \centering
 \includegraphics[width=1.99\columnwidth]{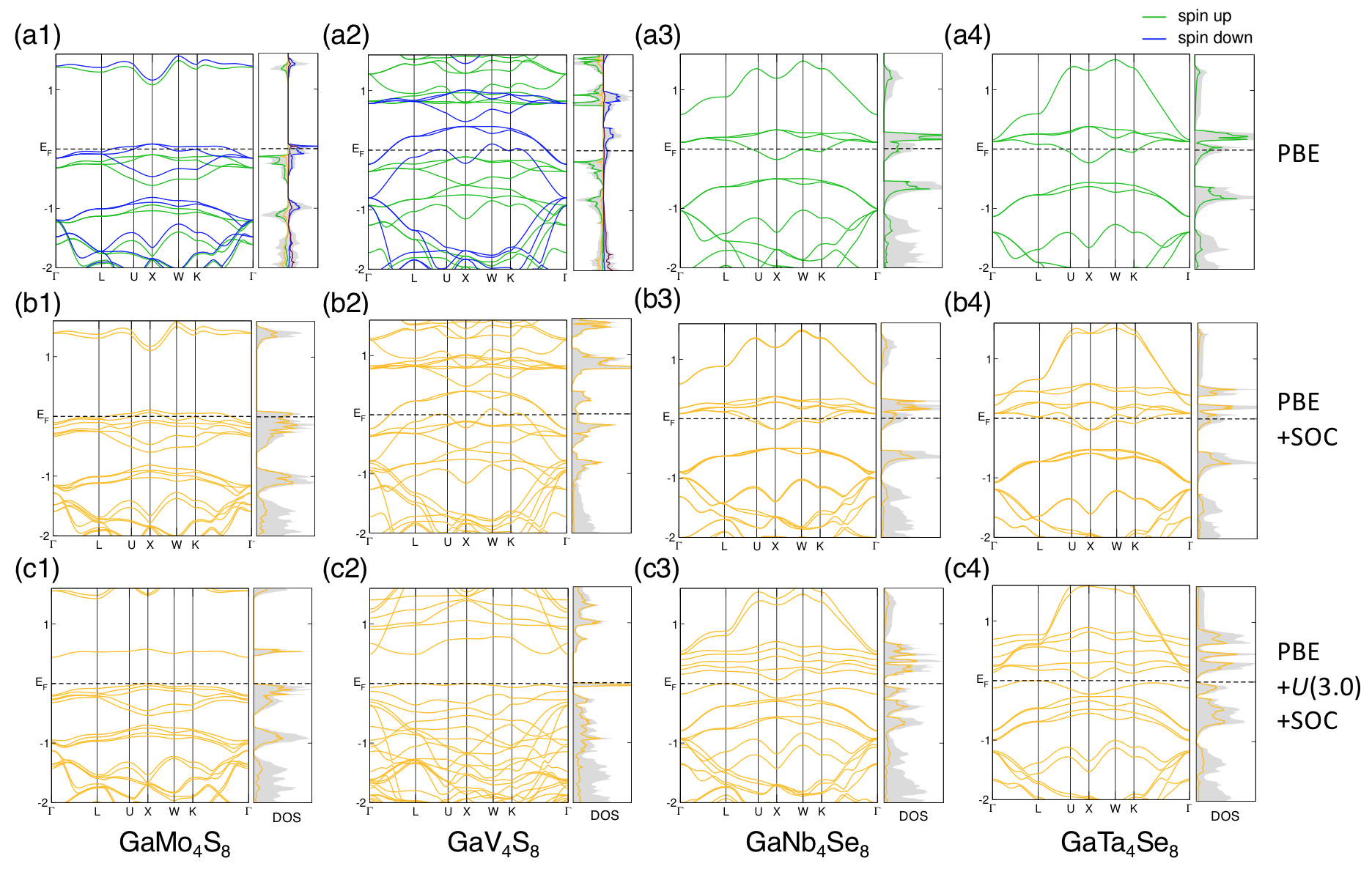}
  \caption{DFT-PBE ground state band structures and \rw{projected} density of states (DOS) of the  cubic lacunar spinels within a primitive cell. The Fermi level ($E_F$) is indicated 
  by a broken line. The gray shaded areas in the DOS panels correspond to the total electronic density of states. \rw{The second and third rows show results with SOC included, as indicated in the rightmost column. The orange curve in the DOSs represent the contribution from the transition metal cluster.}}
  \label{fig:fig5}
\end{figure*}

In several cases, we are not able to stabilize some of the magnetic 
configurations for vanadium compounds (indicated by `--' in \autoref{table:mag_energy_cubic}).
For example, LDA only converges to non-magnetic configurations, 
and PBE$+U=2.0$\,eV cannot stabilize the state with 1\,$\mu_B$ per cluster. 
In cases where both states can be stabilized, however, the more strongly 
spin-polarized state is always significantly more stable than the other 
two configurations (\autoref{table:mag_energy_cubic}). 
We also observe a trend that the highly polarized state is more favored with 
larger on-site Coulomb interactions or with higher level DFT functionals.
In addition, the $\mu = 1\,\mu_B$ state is usually energetically closer  
to the non-magnetic state than the highly spin-polarized state. 
These ground state magnetic configurations are also sensitive to the $V_4$ 
cluster volume, which we show varies with different levels of DFT functional 
(\autoref{fig:fig4}).
A larger $V_4$ cluster usually supports a higher magnetic moment, 
while a smaller volume leads to reduced or quenched moments. 
Our findings show that local structure and magnetic moments are correlated with each 
other and should be assessed carefully because both depend on the choice 
of exchange-correlation functional.
\rw{A recent study utilizing dynamical mean-field theory simulations showed similar results, 
where the significance of electron correlations in describing the MO Mott physics and structural properties of \gvs\, is also reported}\cite{kim2018molecular}.

\rw{There is also evidence that local cluster distortions still exist above the Jahn-Teller temperature\cite{wang2015polar}, 
and that the symmetry-broken V$_4$ cluster could lead to different magnetic configurations that are in better agreement with experimental results\cite{zhang2017magnetic}. Why this occurs is attributed to the physics of the distorted phase described in \autoref{sec:R3m}. To that end, we suggest high-resolution detection methods (e.g. pair distribution function) be used to probe the local structures of cubic phase lacunar spinels.}

Last, the niobium and tantalum compounds are always non-magnetic in our calculations, 
regardless of the initial magnetic configuration or choice of 
DFT functional. 
This may be a consequence of strong but geometrically frustrated antiferromagnetic interactions in the cubic Nb and Ta clusters\cite{pocha2005crystal} or 
due to a reduction in the on-site Hund's interactions, 
which drives moment formation, from the 
greater hybridization from the extended $4d$ and $5d$ orbitals.


\subsubsection{Electronic structures}

We next use the relaxed cubic crystal structure and ground state magnetic configuration of each compound and examine the electronic structures (\autoref{fig:fig5}). 
According to the idealized charge distribution in Ga$^{3+}$[M$_{4}$X$_{4}$]$^{5+}$X$^{2-}_{4}$,
the number of electrons per V$_4$, Nb$_4$, and Ta$_4$ cluster is 7
(since they are in the same column of the periodic table) 
while there are 11 electrons for a Mo$_4$ cluster; 
these electrons fill the cluster orbitals
depicted in \autoref{fig:fig2}.

\begin{figure}[t]
  \centering
 \includegraphics[width=0.9\linewidth]{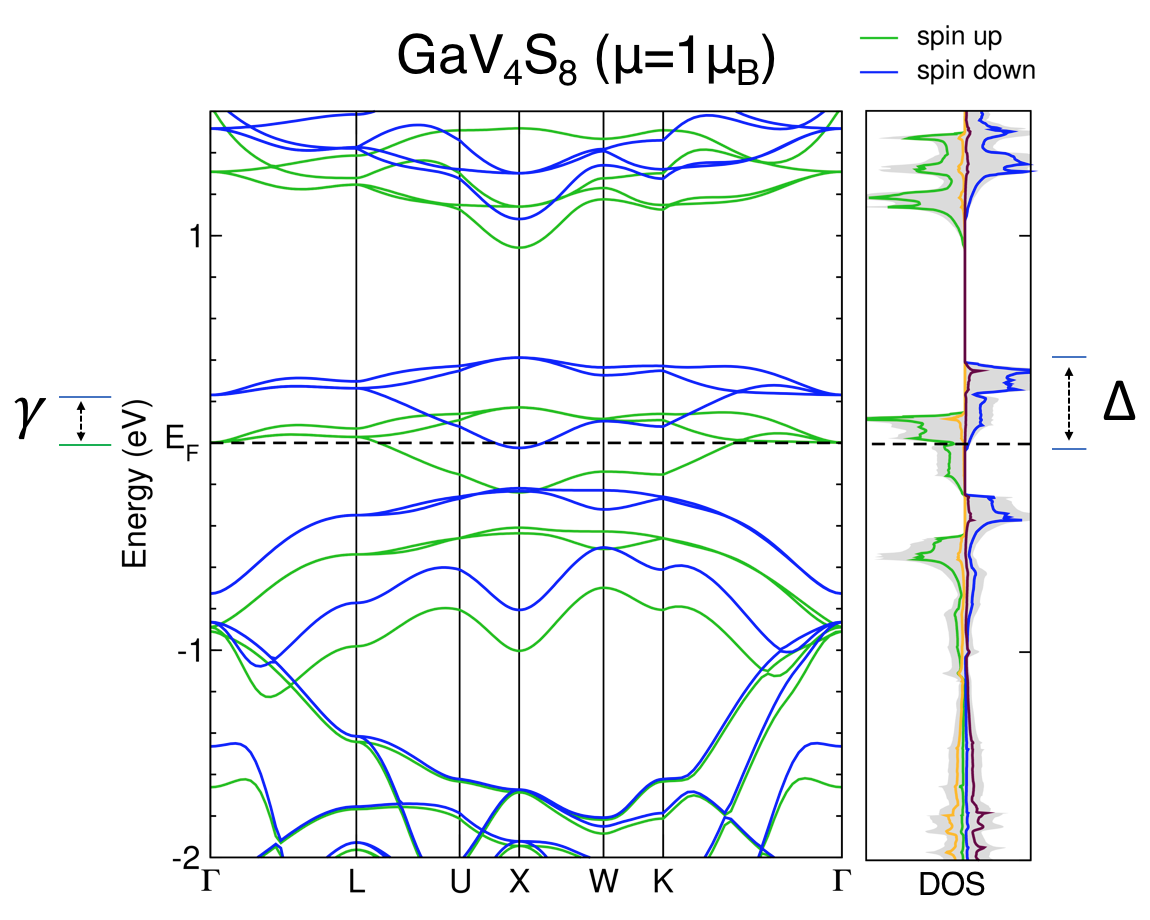}
  \caption{DFT-PBE band structure and DOS of metastable cubic \gvs{} with 1\,$\mu_B$ per formula unit. 
We define $\gamma$ as the exchange splitting between different spin channels and $\Delta$ as the splitting of the three valence bands at the X-point. Here, those bands are located 
approximately at $E_F$ and $E_F+0.3$\,eV.}
  \label{fig:fig6}
\end{figure}

\rw{We selectively show the electronic structures of \gvs, \gms, \gnse, and \gtse\, in \autoref{fig:fig5} because their S/Se counterpart compounds with the same transition-metal cluster exhibit similar band properties.}
From our PBE-DFT band structures and projected DOSs for the 
molybdenum, niobium, and tantalum compounds \rw{(\autoref{fig:fig5} a[1,3,4])},
we find six valence bands mainly consisting of transition-metal $d$-orbital character 
with relatively small contribution from the anion $p$ orbitals.
The triply degenerate band ($t_2$ MO symmetry) is higher in energy than the 
doubly ($e$ MO) and singly ($a_1$ MO) degenerate bands.
The band degeneracy and ordering agree well with the cluster MO descriptions 
of the low-energy electronic structure in these compounds.

The DFT ground state electronic structures of the \rw{cubic phase} vanadium compounds, however, are significantly different from the other chalcogenides in the lacunar spinel 
family \rw{(\autoref{fig:fig5} a2)}.
All six valence bands in the spin-up channel (green bands) are fully occupied, 
while only the lower part of the spin-down channel is partially occupied.
The triply degenerate spin-down bands are shifted $\approx$1\,eV above the Fermi level.
Interestingly, the metastable magnetic state with 1\,$\mu_B$ per cluster
exhibits band dispersions that are more similar to the rest of the family (\autoref{fig:fig6})
and the magnetic moment of 1\,$\mu_B$ agrees better with experimental results.
It remains unknown whether this DFT ground state in the cubic phase is 
stable and experimentally accessible; further low-temperature 
neutron-based scattering measurements, for example, could be used to 
probe the existence of this spin configuration.

We next quantitatively assess \rw{the impact of different $V_{xc}$ as well as on-site Coulomb interactions on} the electronic structures by defining 
two parameters,  $\gamma$ and $\Delta$, as shown in \autoref{fig:fig6}, 
which describe the  key features in the band structure.
$\gamma$ corresponds to the energy difference between different spin-channels 
of the triply degenerate valence band at the $\Gamma$ point.
$\Delta$ quantifies the magnitude of the splitting among the triply-degenerate 
minority-spin bands at the X point, $k=(1/2, 0, 1/2)$, near $E_F$.
The values of $\gamma$ and $\Delta$ for the chalcogenide 
lacunar spinels at different levels of DFT theory are tabulated in  \autoref{tab:band_splitting}.
For the non-magnetic Nb and Ta compounds, we only report $\Delta$.

\begin{table*}
\begin{ruledtabular}
\centering
\caption{\label{tab:band_splitting}Electronic band splitting of the triply degenerate bands in the cubic lacunar spinels at different levels of DFT. $\gamma$ quantifies the splitting between the two spin channels. $\Delta$ is the value of the band splitting among the triply degenerate bands at the  \revsecond{X point in momentum space} near $E_F$. For the vanadium compounds, these values are tabulated for different spin-magnetic moment states separately. An `--' indicates that the state was not stable.}
\begin{tabular}{lcccccccc}%
compound &  & LDA & PBEsol & PBE & PBE$+U(1.0)$ & PBE$+U(2.0)$ & SCAN & HSE06 \\
\hline
\gms & $\gamma$ & 0.09	& 0.15	& 0.16	& 0.22	& 0.28	& 0.22	& 0.56\\
& $\Delta$ & 0.69	& 0.65	& 0.55	& 0.55	& 0.55	& 0.59	& 0.67\\
\hline
\gmse & $\gamma$ & 0.11	& 0.14	& 0.16	& 0.22	& 0.29	& 0.22	& 0.52\\
& $\Delta$ & 0.50	& 0.48	& 0.38	& 0.39	& 0.40	& 0.44	& 0.52\\
\hline
\gvs  & $\gamma$ & -- &	--	& 1.13 & 	1.58	 & 2.01 &	 1.76 & 3.1 \\
(5\,$\mu_B$) & $\Delta$ & --	& -- & 	0.48	 & 0.44	& 0.33	& 0.43 & 0.46 \\
\hline
\gvs  & $\gamma$ & -- &	0.12 & 	0.23	& 0.32 &	 -- & 0.37 & 0.95\\
 (1\,$\mu_B$) & $\Delta$ & -- & 0.52	& 0.41 & 	0.40 &	-- &	 0.43	& 0.55 \\
\hline
\gvse  & $\gamma$ & -- &	1.07 & 	1.17 & 1.65 & 2.60 & 1.82 & 3.2 \\
(5\,$\mu_B$) & $\Delta$ & -- & 0.42 & 0.36 & 0.34 & 0.20 & 0.35 & 0.51 \\
\hline
\gvse  & $\gamma$ & -- & 0.20 & 0.23 & 0.34 & -- & 0.40 & 0.81 \\
(1\,$\mu_B$) & $\Delta$ & --	& 0.37 & 0.30 & 0.28 & -- & 0.31 & 0.41 \\
\hline
\gns & $\Delta$ & 0.88 & 0.82 & 0.69 & 0.71 &	0.71 &  0.73 & 0.86 \\
\hline 
\gnse & $\Delta$ & 0.63 & 0.59 & 0.50 & 0.52 &	 0.53 & 0.55 & 0.66 \\
\hline
\gtse & $\Delta$ & 0.74 & 0.71	& 0.60 & 0.63 & 0.63 & 0.65 & 0.69
\end{tabular}
\end{ruledtabular}
\end{table*}

All of the cubic phase lacunar spinels are metallic from band theory \rw{without considering spin-orbit interactions}. 
\autoref{fig:fig5} shows that the Fermi level, $E_F$, is always located 
within the valence bands, 
regardless of the magnetic configuration or DFT functional.
Specifically, \rw{the cubic phase V and Mo compounds} are predicted to be half-metals 
as only one spin channels cross the Fermi level whereas the other spin-channel is fully gapped. 
In either group VB or VIB transition metal compounds, there is an odd number of electrons in three degenerate bands (\autoref{fig:fig2}).
For the low spin-polarized states with 1\,$\mu_B$ magnetic moment per formula unit, we then find that the Fermi level 
crosses this set of triply degenerate bands and metallicity is protected by 
the $F\bar{4}3m$ crystal symmetry.
Here the splitting of these triply degenerate valence bands throughout 
the Brillouin zone is quite small; although the $\Delta$ value is functional dependent, it does not exceed 0.7\,eV.
There is also a small trend of increasing splitting between different spin channels ($\gamma$) with higher levels DFT functionals.
We attribute this to the 
more accurate exchange interactions captured with the more advanced functionals.

The flat valence bands derived from these cluster orbitals 
lead to large effective masses, 
and these electrons should be highly localized in real space.
This is in agreement with the fact that the transition-metal clusters are 
far from each other within the unit cell,
and the electrons are highly localized within the cluster.
One of the possible conduction mechanisms for the lacunar spinels is through  variable-range hopping (VRH) \cite{sahoo1993evidence}.
It is for the same reason that these compounds have been called ``Mott insulators''\cite{pocha2000electronic}.
%

For the highly-polarized magnetic state in the vanadium compounds, 
$\gamma$ is much larger than $\Delta$, which makes 
it different from the rest of the family.
In this case, the $\Delta$ term may not be that important since the triply degenerate band is no longer the highest occupied band.
The two bands crossing the Fermi level are the 
$a_1$ and $e$ orbitals in the spin-down channel.
It is therefore possible to obtain a semiconducting state by shifting 
the $e$-symmetry orbitals to higher energy and fully occupying the 
$a_1$ orbital.
Indeed, we find such a state in \gvse\ using the SCAN functional, 
where the band gap is approximately 60\,meV. 
Whether this highly-polarized state is experimentally accessible, 
however, remains unclear.

\rw{%
We next report results with SOC included in our simulations. 
The band structures and DOSs with the PBE functional are shown in \autoref{fig:fig5} b[1-4]. 
Orbital degeneracy is partly broken compared with the non-SOC band structures. 
The broken symmetry here is vital for reproducing a semiconducting state 
since it enables further orbital splitting by increasing electron-electron interactions.
\autoref{fig:fig5} c[1-4] show the electronic structures with PBE$+$SOC and a $U$ value of 3.0\,eV, where all four compounds exhibit a small but finite band gap.
It is interesting to note that both SOC and on-site Coulomb interactions are necessary 
in order to produce a semiconducting cubic phase for all compounds studied.
Intuitively, SOC serves the purpose of symmetry-breaking in the highly-symmetric cubic phase 
while on-site Coulomb interactions localize electrons and increase repulsion between bands,
which eventually lead to a semiconducting state in the cubic lacunar spinels.
Although the electron-correlation effect (modeled by the Hubbard $U$) is typically considered more important in 3$d$ transition metals,
spin-orbit interactions are more significant in 5$d$ transition metals.
Indeed, the lacunar spinel compounds investigated, which include transition metals from 
the 3$d$, 4$d$, and 5$d$ rows, exhibit similar yet non-identical behaviors.
This behavior could be the outcome of competing SOC and on-site Coulomb interactions within these transition-metal cluster systems.}
\rw{It has been shown that spin-orbit coupling effect within the lacunar spinel system could lead to exciting physics (e.g. spin-orbital entangled molecular $j_\mathrm{eff}$ states).}\cite{kim2014spin,jeong2017direct}

Our findings \rw{in the cubic phase lacunar spinels} indicate that
different DFT functionals, \rw{as well as various internal electron-electron, spin-orbital interactions}, can lead to qualitatively different 
interpretations of their electronic and magnetic properties.
Therefore, extra care in the exchange-correlational 
functional selection should be taken before 
pursuing extensive DFT simulations on this family. 


\subsection{$R3m$ distorted phase}\label{sec:R3m}


\subsubsection{Lattice parameters}

In this section, we investigate the DFT functional dependency of properties in the distorted rhombohedral phase.
Since only molybdenum and vanadium compounds are reported to exhibit Jahn-Teller-type structural distortions,
\rw{we benchmark the $V_{xc}$ performance in predicting lattice parameters against available experimental data of \gvs, \gvse, \gms, and \gmse.}
%
In all cases, we use a ferromagnetic spin configuration with 
1$\mu_B$ magnetic moment per unit cell in our structural 
relaxations; see \autoref{sec:mag_R3m} for a detailed discussion 
of the magnetic moment configurations. 

\begin{figure}
\centering
\includegraphics[width=0.95\linewidth]{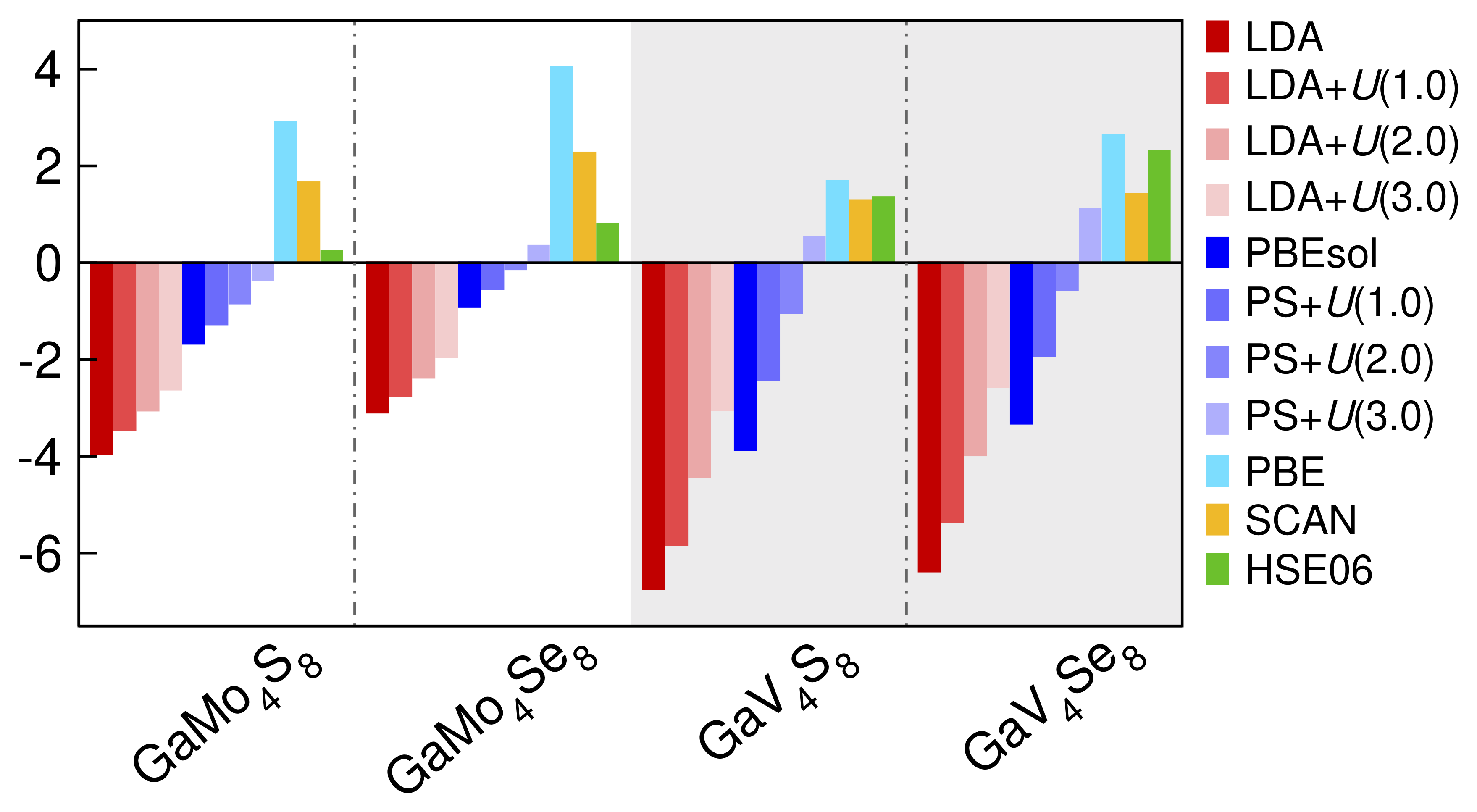}
\caption{Relative error of the rhombohedral unit cell volume at different  levels of DFT. PS is an abbreviation for the PBEsol functional and the number in parentheses is the value of the on-site Coulomb interaction used in the \rw{LDA/GGA$+U$ method}.}
\label{fig:fig7}
\end{figure}

The Jahn-Teller structural phase transition reduces 
the crystal symmetry from space group $F\bar{4}3m$ to $R3m$ and occurs with a change in unit cell volume.
The relative error of the fully relaxed unit cell volumes for the molybdenum and vanadium compounds are shown in \autoref{fig:fig7}.
Here, we observe a similar trend as found in the cubic phase.
The LDA and PBEsol functionals underestimate the ground state lattice volume, while LDA shows larger deviations from the experimental data.
%
%
\rw{Moreover, structural relaxations of the rhombohedral 
phases of \gvs\, and \gms\, with LDA converge to 
non-magnetic cubic structures,
regardless of the initial magnetic moment configurations.
LDA is able to stabilize a ferromagnetic configuration in the rhombohedral phase only with on-site Coulomb interactions (LDA+$U$).}

PBE overestimates the lattice constants of all four compounds.
With increasing value of the on-site Coulomb interactions, the lattice parameters also increase slightly.
In the rhombohedral phase, the cell volume of vanadium compounds is not as sensitive to the Hubbard-$U$ value as in the cubic phase, presumably because the electronic 
structure is semiconducting in the $R3m$ symmetry.
SCAN and HSE06 functional again perform quite well with 
regards to the lattice parameters with less than 2\%  error.
%


\subsubsection{Internal degrees of freedom\label{sec:R3m_dof}}

\begin{figure}[h]
\centering
\includegraphics[width=0.95\linewidth]{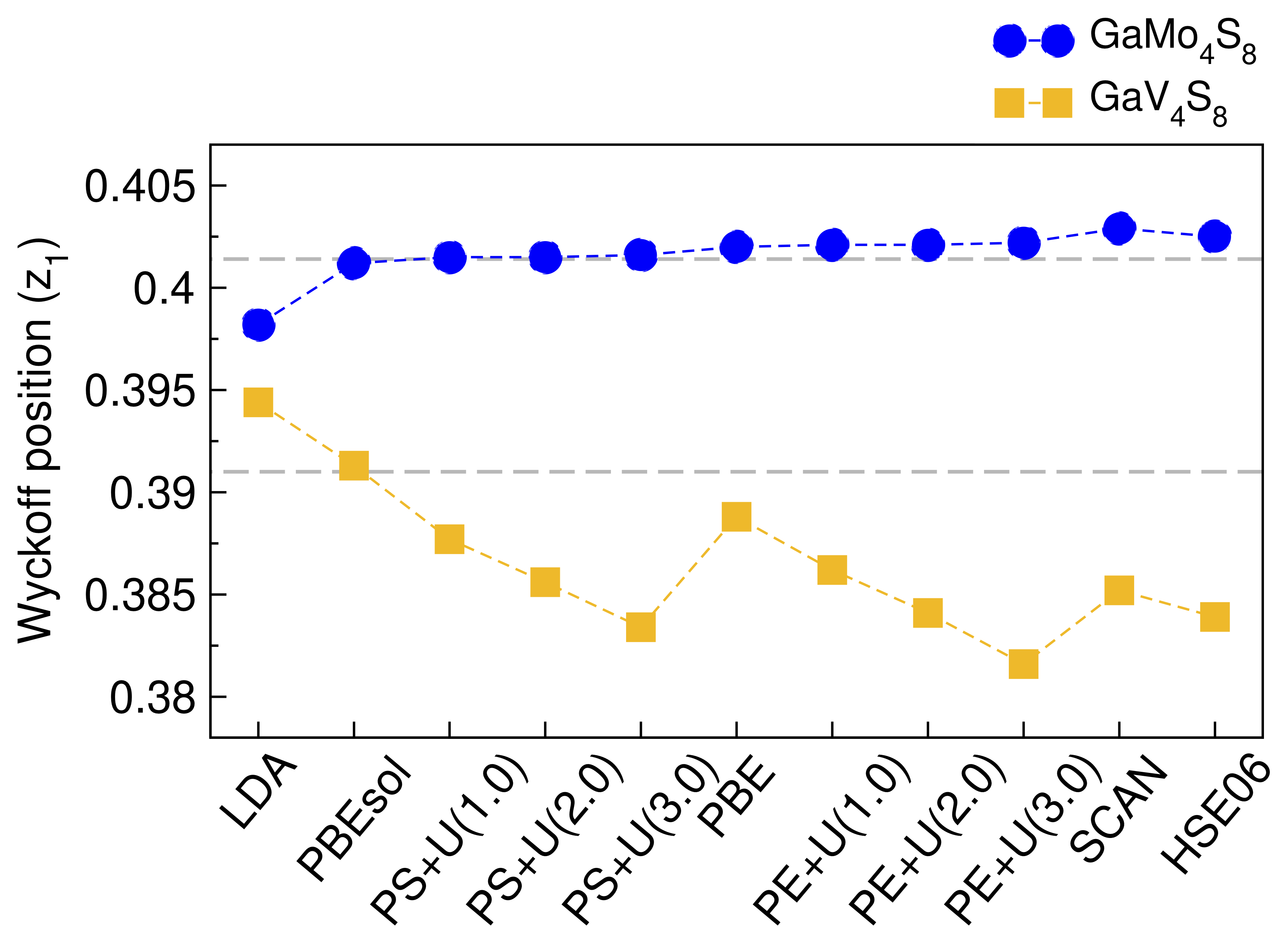}
\caption{The $3a$ ($z_1$) Wyckoff position in rhombohedral \gms\, and \gvs\, at different $V_{xc}$. The gray dashed lines correspond to the experimental values. \revsecond{PS is an abbreviation for the PBEsol functional and the number in parenthesis is the value of the on-site Coulomb interaction used in the GGA$+U$ method.}}
\label{fig:fig8}
\end{figure}

The occupied Wyckoff sites of the transition metals also split upon the transition into the rhombohedral phase, 
leading to one apical site [M$_1$ in \autoref{fig:fig1}(b)] along the $C_{3v}$ distortion axis 
and three basal atoms [M$_2$, M$_3$, M$_4$ in \autoref{fig:fig1}(b)] forming a plane perpendicular to the $C_{3v}$ axis.
The Wyckoff positions of the transition metals in \gms\ and \gvs\ with $R3m$ symmetry (space group no.\ 160) after structural relaxation with different exchange-correlation functionals are tabulated in \autoref{tab:wyckoff_positions}. 
%
%
The selenide compounds show similar functional dependencies and are not shown here. 
In general, the changes in Wyckoff positions with respect to functional  are quite small. 
However, we find that the $z_1$ value in \gvs\, has a significantly higher functional dependency over that in \gms\,(\autoref{fig:fig8}).
Both increasing the value $U$ as well as going to higher levels of exchange-correlation functionals  favor larger structural distortions in \gvs, i.e.,  keeping the apical V atom far away from the center of the tetrahedral transition-metal cluster. The $z_1$ Wyckoff position of the Mo atoms is also largely insensitive to the choice of the DFT functional, possibly owing to the reversed distortion in \gms, where steric effects might prohibit further distortion.

\begin{table}
\begin{ruledtabular}
\centering
\caption{\label{tab:wyckoff_positions} Wyckoff positions of the transition metals in rhombohedral \gms\, and \gvs\, after structural relaxation with different DFT functionals. The $z$ value of the $3a$ and $9b$ sites in space group no.\ 160 are labeled $z_1$, $z_2$, respectively. \revsecond{PS is an abbreviation for the PBEsol functional and the number in parenthesis is the value of the on-site Coulomb interaction used in the GGA$+U$ method.}
}
\begin{tabular}{lccc}%
 & \multicolumn{3}{c}{\gms}  \\
\cline{2-4}
 & $3a$ ($z_1$) & $9b$ ($x$) & $9b$ ($z_2$) \\
\hline
experimental\cite{powell2007cation} & 0.4014 & 0.1956 & 0.2023 \\
\hline
LDA & 0.3982 & 0.1960 & 0.2022 \\
\hline
PBEsol & 0.4012 & 0.1951 & 0.2012 \\
\hline
PS$ + U(1.0)$ & 0.4014 & 0.1950 & 0.2011\\
PS$ + U(2.0)$ & 0.4015 & 0.1949 & 0.2010\\
PS$ + U(3.0)$ & 0.4016 & 0.1948 & 0.2010\\
\hline
PBE & 0.4020 & 0.1956 & 0.2009 \\
\hline
SCAN & 0.4029 & 0.1962 & 0.2006\\
\hline
HSE06 & 0.4025 & 0.1959 & 0.2007 \\
\hline
\hline
 & \multicolumn{3}{c}{\gvs}  \\
\cline{2-4}
 & $3a$ ($z_1$) & $9b$ ($x$) & $9b$ ($z_2$) \\
 \hline
 experimental\cite{powell2007cation} & 0.3910 & 0.1937 & 0.2013\\
\hline
LDA &  0.3944 & 0.1946 & 0.1998\\
\hline
PBEsol & 0.3913 & 0.1969 & 0.2005 \\
\hline
PS$ + U(1.0)$ & 0.3877 & 0.1966 & 0.2005\\
PS$ + U(2.0)$ & 0.3856 & 0.1958 & 0.2005\\
PS$ + U(3.0)$ & 0.3834 & 0.1951 & 0.2008\\
\hline
PBE & 0.3888 & 0.1972 & 0.2005 \\
\hline
SCAN & 0.3852 & 0.1963 & 0.2004\\
\hline
HSE06 & 0.3839  & 0.1956 & 0.2005
\end{tabular}
\end{ruledtabular}
\end{table}

After the structural phase transition, both the rhombohedral angle $\alpha_{rh}$ and bond angle $\theta_m$ in the transition-metal cluster
diverge from the cubic 60$^{\circ}$, leading to a greater number of internal degrees of freedom in the distorted phase.
The latter is correlated with the change in 
occupied Wyckoff sites of the transition metals .
We record these internal bond angles of the four lacunar spinels after structural relaxation using different DFT functionals,
the results are shown in \autoref{tab:internal_coord}.

\begin{table*}[tbh]
\begin{ruledtabular}
\centering
\caption{The unit cell rhombohedral angle $\alpha_{rh}$ (in degrees) for the $R3m$ phases and the corresponding apical bond angle $\theta_m$ (in degrees) for the transition-metal cluster at different levels of DFT functional. \revsecond{PS is an abbreviation for the PBEsol functional and the number in parenthesis is the value of the on-site Coulomb interaction used in the GGA$+U$ method.}\label{tab:internal_coord}}
\begin{tabular}{lcccccccc}%
 & \multicolumn{2}{c}{\gms} & \multicolumn{2}{c}{\gmse} & \multicolumn{2}{c}{\gvs} & \multicolumn{2}{c}{\gvse} \\
\cline{2-3} \cline{4-5} \cline{6-7} \cline{8-9}
	& $\alpha_{rh}$	& $\theta_m$ & $\alpha_{rh}$	 & $\theta_m$ & $\alpha_{rh}$	& $\theta_m$ & $\alpha_{rh}$ & $\theta_m$ \\
\hline
experimental & 60.47 & 61.60 & 60.57 & 61.43 & 59.62 & 58.38 & 59.56 & 57.72\\
\hline
LDA	& 60.00 & 60.00 & 60.78 & 62.76 & 60.00 & 60.00 & 59.55 & 57.71\\
\hline
PBEsol & 60.70 & 62.29 & 60.80 & 62.91 & 59.56 & 57.67 & 59.33 & 56.59 \\
\hline
PS$+U(1.0)$ & 60.75 & 62.50 & 60.81 & 63.00 & 59.28 & 56.39 & 59.20 & 56.00 \\
PS$+U(2.0)$ & 60.76 & 62.59 & 60.81 & 63.06 & 59.16 & 55.91 & 59.09 & 55.60 \\
PS$+U(3.0)$ & 60.77 & 62.66 & 60.81 & 63.12 & 58.99 & 55.26 & 58.87 & 54.75 \\
\hline
PBE & 60.73 & 62.53 & 60.79 & 63.07 & 59.36 & 56.62 & 59.26 & 56.09 \\
\hline
SCAN & 60.76 & 62.80 & 60.84 & 63.27 & 59.21 & 55.72 & 59.12 & 55.47 \\
\hline
HSE06 & 60.77 & 62.74 & 60.79 & 63.15 & 59.05 & 55.38 & 58.94 & 55.04 \\
\end{tabular}
\end{ruledtabular}
\end{table*}

Almost all DFT functionals (except for LDA) predict  similar results for $\alpha_{rh}$ compared with the experimental data, 
but in general they give larger local metal-cluster distortions.
$\theta_m$ values are 1$\sim$2 degrees larger than experimentally reported in the molybdenum compounds, 
whereas the vanadium compounds show a similar but reversed trend in $\theta_m$, i.e.,  1$\sim$2 degrees smaller.
%
This agreement is reasonable, and the difference in internal coordinates compared with experiment
might come from low-resolution experimental characterization \cite{sieberer2007importance}.
It is also possible that the structural phase transition is incomplete at low temperature\cite{francois1991structural}.
We also find a minor trend 
\rw{that higher-level functionals,}
as well as larger Hubbard-$U$ values,
favor larger structural distortions.
Since our DFT simulations are performed at 0\,K, while  lab characterizations are performed at finite temperature,
our results are more likely to capture the correct ground state structure where thermal expansion 
effects are small.

It is interesting to note that the vanadium and molybdenum compounds show reversed structural distortions  across the phase transition.
This can be explained from the valence MO diagram in \autoref{fig:fig2}.
In the cubic phase, there is either 1 electron (\gvs) or 5 electrons (\gms) in the valence $t_2$ orbitals.
Such electronic configurations are Jahn-Teller active, 
whereby a structural distortion accompanied by orbital-degeneracy lifting could further stabilize the system.
Owing to the different electron occupations in the vanadium and molybdenum compounds,
their favored electronic configurations require a reversed ordering of the $a_1$ and $e$ orbitals.
Therefore, the spontaneous structural distortion permits each compound to lift its orbital degeneracy and 
achieve its favored electronic configuration. 

We now summarize the structural benchmark assessment of the cubic and rhombohedral phases. 
We recommend using the GGA$+U$ method for lattice structure relaxations with a Hubbard-$U$ value of 
approximately \rw{2\,eV to 3\,eV}. 
%
\rw{LDA functional should be used with on-site Coulomb interactions for both cubic and rhombohedral phases.}
The SCAN functional is another reasonable choice 
that predicts accurate lattice structures.
Structural relaxations with HSE06 give very accurate lattice constants, but its high computational 
costs may be prohibitive if only trying to obtain reasonable crystal structures. 
%


\subsubsection{Magnetism\label{sec:mag_R3m}}

We find that the DFT ground state of both the molybdenum and vanadium compounds in the distorted $R3m$ structure  
are 
ferromagnetic
with 1\,$\mu_B$ per formula unit.
(Simulation of the complex magnetic phase diagram of lacunar spinels is out of the scope of this work, readers with interest should refer to Refs.\ \onlinecite{franke2018magnetic,widmann2017multiferroic,butykai2017characteristics,ruff2015multiferroicity,zhang2019possible,kezsmarki2015neel,ehlers2016skyrmion,leonov2017skyrmion}.)
In the molybdenum compounds, the magnetic moment is evenly distributed about all four Mo atoms,
which is the same as what we found in the cubic phase.
In the vanadium compounds, however, the apical V atom  along the $C_{3v}$ symmetry axis has a large local magnetic moment.
The other three basal V atoms have relatively smaller moments that are anti-aligned to the apical spin.
This results in  a ferrimagnetic configuration in the V$_4$ cluster, giving a net-magnetic moment of 1\,$\mu_B$ per formula unit.
%
For instance, in the rhombohedral phase of \gvs{} with the PBE functional, the magnetic moment on the apical V atom is 1.3\,$\mu_B$  while the three basal V atoms contribute each $-$0.1\,$\mu_B$. Thus, the net-magnetic moment in one formula unit of \gvs\, is 1\,$\mu_B$.
\rw{Some DFT functionals (e.g.\ PBE$+U(1.0)$) are able to stabilize a ferromagnetic configuration in \gvs\, similar to that of \gms, i.e.\ evenly distributed, but this magnetic configuration is less energetically favorable compared with the ferrimagnetic configuration. We report properties of the rhombohedral phase vanadium compounds using the ferrimagnetic configuration in the remainder of the manuscript.}
\rw{A recent work that used random-phase approximation correctly reproduced the ground state of \gvse\, and explored the coupling between magnetism and structure.}\cite{schueller2019modeling}
%


\subsubsection{Electronic structures}

\rw{We examine the electronic structures of rhombohedral \gvs, \gms, \gnse, and \gtse\, in this section, 
because there is evidence that symmetry-breaking in the transition-metal clusters without distortion of the lattice parameters could lead to different magnetic configurations.\cite{zhang2017magnetic} 
Such small local distortions may also be challenging to detect with low-resolution characterization techniques; for that reason, 
we hypothesize that the niobium and tantalum-based lacunar spinels could also exhibit a distorted rhombohedral phase.}
\rw{Therefore, we slightly distort the cubic niobium and tantalum lacunar spinel structures along the symmetry-breaking pathway, 
and use these geometries as the initial structure for structural relaxations. 
The structural relaxation settings using different DFT functionals are similar to those used for the cubic phase.}

\begin{figure*}
\centering
\includegraphics[width=1.99\columnwidth]{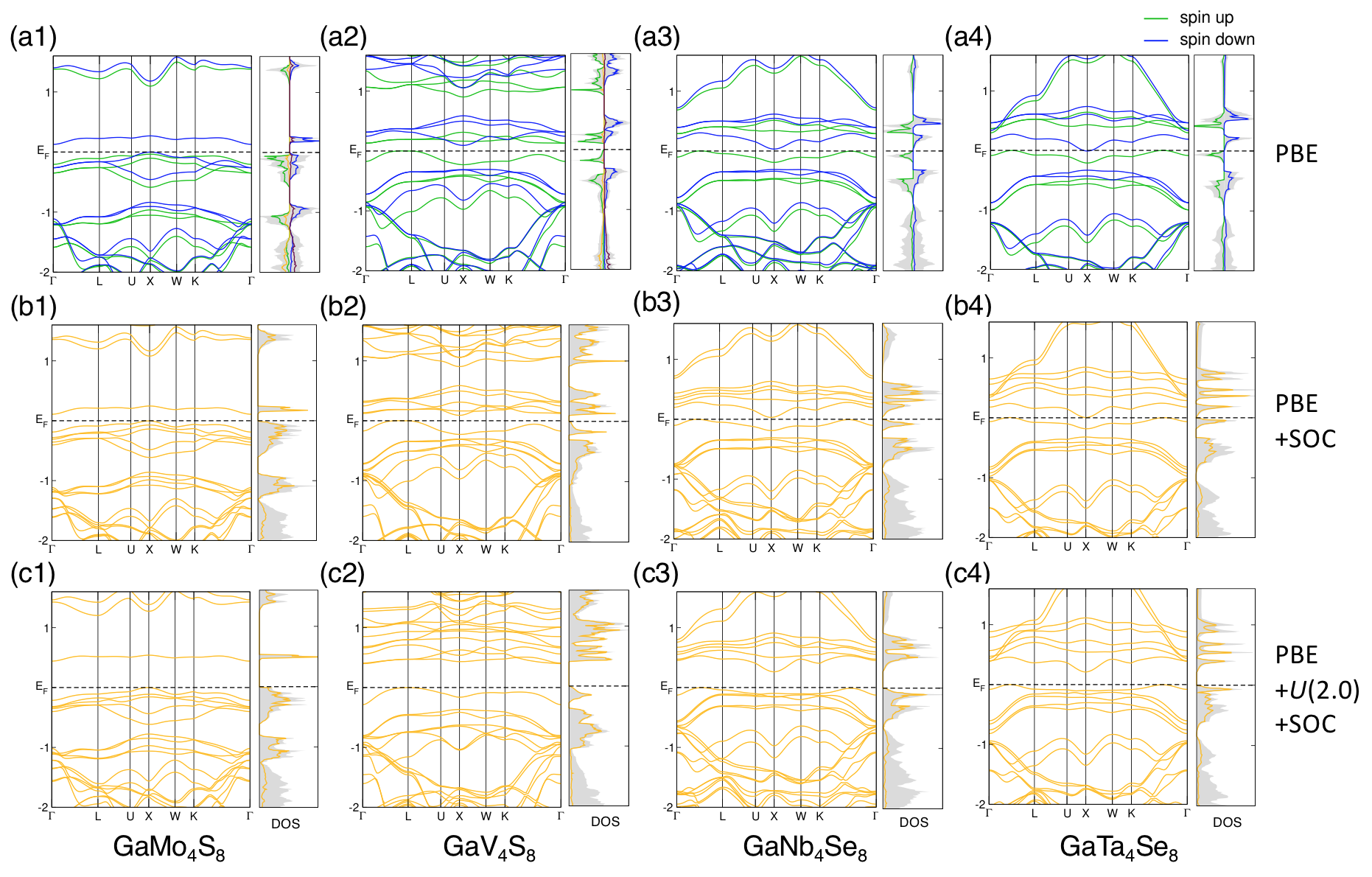}
\caption{DFT-PBE ground state band structures and density of states (DOS) of the rhombohedral $R3m$ lacunar spinels. \rw{The second and third rows show simulation results with SOC. Color representation is the same as that in \autoref{fig:fig5}}.}
\label{fig:fig9}
\end{figure*}

%
\rw{\autoref{fig:fig9} a[1-4] presents the electronic band structures and projected DOSs of these four compounds.}
The triply degenerate valence bands in the cubic 
phase split into two sets of orbitals with $a_1$ and $e$ symmetry.
In the molybdenum compounds, the minority spin $a_1$ 
orbital shifts to higher energy, above the Fermi level,
\rw{such that five valence electrons occupy the 
three majority spin orbitals and the minority $e$ orbitals.}
\rw{The vanadium, niobium, and tantalum compounds} exhibit different orbital  occupations and structural distortions; 
the $a_1$ orbital is further stabilized to 
lower energy relative to the other five orbitals,
and the only one valence electron occupies the $a_1$ orbital.
Remarkably, we find that the PBE functional is able to open up a small band gap without any on-site Coulomb interactions in the distorted phase \gvs\, and \gms. 
\rw{For \gnse\, and \gtse, the lowest conduction band (minority $a_1$) barely touches the Fermi level.
However, LDA predicts an unreasonable metallic ground state for these compounds.}
This finding indicates that the structural distortion alone is sufficient to lift the orbital degeneracy and 
open a semiconducting gap \rw{without strong electron-correlation effect}---apparently the additional electron 
density gradient in $V_{xc}$ through the enhancement factor provides an improved description.
In other words, the rhombohedral lacunar spinels  may not be strictly described as ``Mott'' insulators.
Our findings confirm the importance of local structural distortions on electronic structures in lacunar spinels\cite{sieberer2007importance},

\begin{figure}
  \center
  \vspace{-1mm}
 \includegraphics[width=1.0\linewidth]{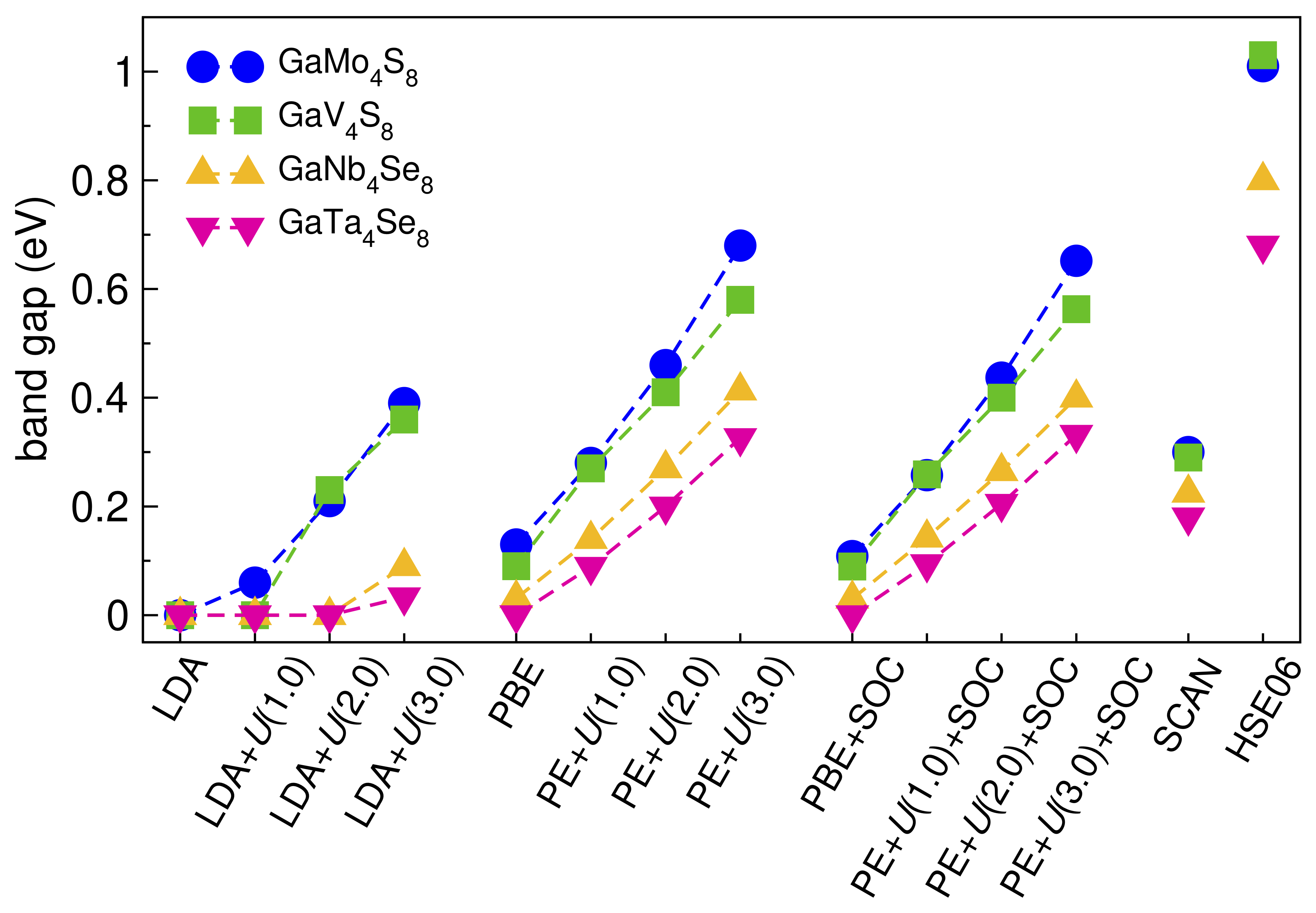}
  \caption{DFT-functional dependence of the electronic band gaps for lacunar spinels in the rhombohedral $R3m$ structure.}
  \label{fig:fig10}
\end{figure}

\begin{SCfigure*}
  \centering
 \includegraphics[width=1.35\columnwidth]{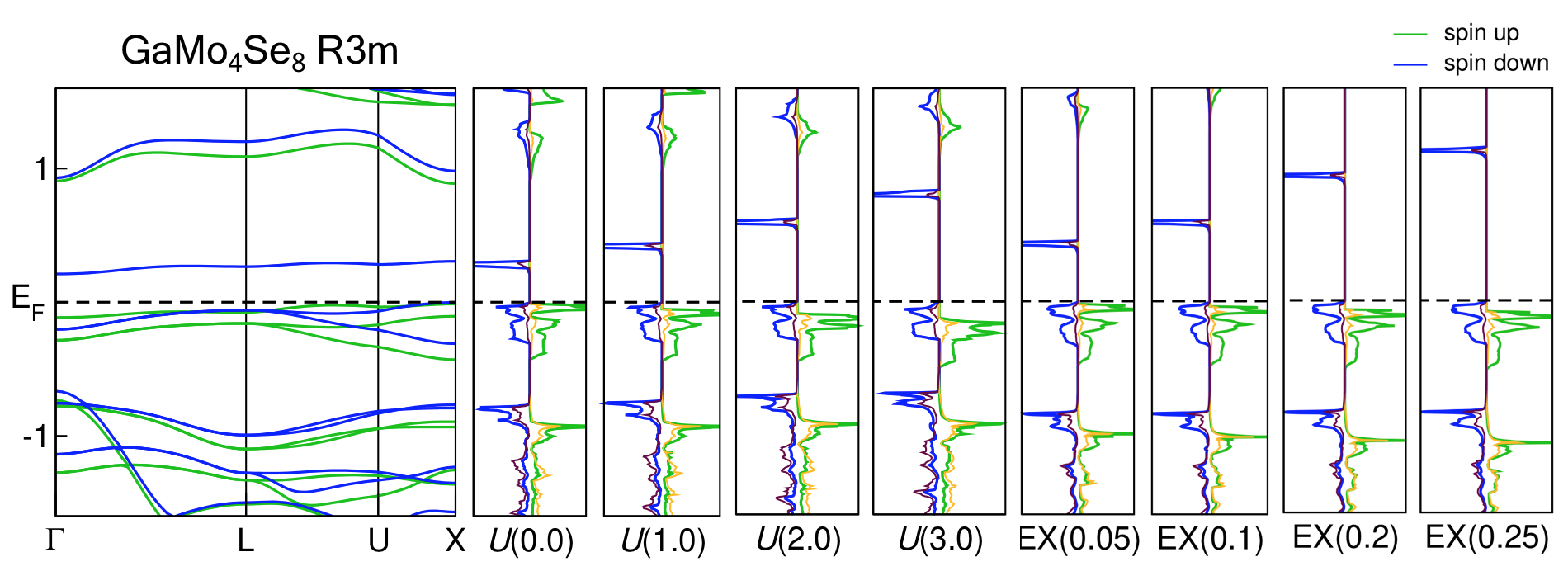}
  \caption{\label{fig:fig11} Effect of on-site Coulomb interactions ($U$ values of 0.0, 1.0, 2.0, and 3.0\,eV) and the amount of exact exchange interactions in the hybrid HSE functional (EX values of 0.05, 0.1, 0.2, and 0.25) on the electronic structures of rhombohedral $R3m$ \gmse. $\mathrm{EX}=0.25$ corresponds to the standard amount of exact exchange in HSE06. The band structure panel on the left is obtained using the PBE functional.}  
\end{SCfigure*}

\rw{Meanwhile,} different GGA functionals qualitatively describe the rhombohedral electronic structures 
differently.
For instance in \gvs, PBEsol predicts a metallic state, while PBE opens a small gap of 0.09\,eV.
In \gms, PBEsol gives a very small band gap of 0.02\,eV  
while PBE predicts a band gap of 0.13\,eV, which is much closer to the experimentally estimated value of 0.2\,eV.
Therefore, DFT-GGA simulations on this family of compounds should be performed with extra caution with 
attention focused on the role of the enhancement 
factor in reducing the self-interaction error.\cite{PhysRevLett.82.2544}
\rw{We recommended that when using the PBEsol functional to simulate the electronic structures of the lacunar spinels, a slightly larger ($\sim$1\,eV) Hubbard $U$ value is used than that for PBE.}
%

\rw{\autoref{fig:fig9} b[1-4] shows the electronic structures with the PBE functional and SOC. 
The effect of including SOC is similar to that in the cubic phase;  
orbital degeneracy is lifted, 
but the overall band structures remain similar. 
\autoref{fig:fig9} c[1-4] shows the effect of now adding a 
Hubbard on-site Coulomb interaction of 2.0\,eV. 
All four compounds now exhibit a clear band gap, 
where the conduction band is pushed to higher energy  owing to stronger electron-electron interactions.
Interestingly, we find that SOC does not seem to play a decisive role in predicting reasonable electronic structures in the rhombohedral phase, 
whereas the GGA functional alone could predict qualitatively correct behavior.
It is possible that SOC plays a less significant role here since crystal symmetry is already broken in the rhombohedral phase, unlike in the highly symmetric cubic phase.
Our findings here support our previous hypothesis about the roles of SOC and electron-correlation effect in predicting semiconducting phases.}

\autoref{fig:fig10} shows the different band gaps predicted using different DFT functionals.
%
\rw{The decreasing band gap in the V-Nb-Ta series from 3$d$ to 5$d$ agrees well with our physical intuition,
where electron-correlation effects are expected to decrease.
The reason why molybdenum compounds show large band gaps might be caused by different orbital occupations 
-- more valence electrons lead to larger orbital repulsion, 
which pushes the conduction band to a higher energy level, 
leading to a larger band gap.}
We also observe a higher functional rather than compositional dependency on the band gap.
With an increasing Hubbard-$U$ value, the band gap increases monotonically.
Since a larger on-site Coulomb interaction effectively increases the repulsion between bands with the same spin,
it results in a larger gap between the highest occupied  and lowest unoccupied bands.
%
%
%

We also observe an interesting similarity in \autoref{fig:fig8} and \autoref{fig:fig10}, 
where the functional dependency of the Wyckoff position $3a$ ($z_1$) in \gvs\, is similar to the trend in the band gap.
\rw{A larger structural distortion in \gvs\, also leads to a higher electronic band gap.
\gms\, however, does not exhibit such correlated properties.}
The distinct behaviors of the Mo and V compounds indicate rather different relationships between the structural distortion and the ground state electronic structures. 
Niobium and tantalum compounds have relatively smaller band gaps compared with vanadium and molybdenum ones, 
this is consistent with experimental estimation of band gaps.
%
\rw{It is also clear that including SOC has a negligible effect on ground state band gap in all four compounds studied here.}
%

\rw{SCAN predicts band gaps close to the experimentally suggested 0.2$\pm$0.1\,eV value\cite{cario2010electric}, whereas 
HSE06 finds approximately a 1.0\,eV band gap for the vanadium and molybdenum chalcogenides, and around 0.7\,eV for the niobium and tantalum compounds.
Since the hybrid functionals partially correct the self-interaction problem in DFT, it is expected to predict 
more accurate band gaps than lower rung functionals.
\revsecond{The larger portion of non-local and range-separated exact exchange interactions included in HSE06, however, might also destroy the balance within the transition-metal cluster, causing the large deviations in the band gaps of the lacunar spinels.}
It has also been reported that the ferromagnetic ground state is determined by the symmetric exchange interactions \cite{zhang2017magnetic}, 
which could possibly explain the different behaviors of HSE06 from lower-level functionals.
More careful experimental characterization of the distorted phase band gaps is required to have a better understanding of which $V_{xc}$ performs the best.}

We next examine the effect of the on-site Coulomb interactions and exact exchange interactions 
on the electronic structures of rhombohedral \gmse{} (\autoref{fig:fig11}). 
The band structure and DOS of \gmse{} using PBE functional with Hubbard-$U$ values of 0.0, 1.0, 2.0, and 3.0\,eV 
are shown in the first five panels.
The four panels starting from the right of 
\autoref{fig:fig11} correspond to the DOS obtained using the HSE06 
functional with different portions of exact exchange included as indicated in parenthesis.
In general, we observe very similar DOS for the occupied bands. The three valence bands in the spin-up channel are slightly shifted to lower energy relative to the Fermi level, $E_F$ with either larger $U$ values or larger amounts of exact exchange.
The orbitals beneath these valence orbitals, 
approximately located at $-1$\,eV, are always lower in energy in our HSE06 calculations.
Because the HSE06 functional treats all orbitals on 
the same footing, these lower energy orbitals are also 
`corrected' in a self-consistent manner, 
whereas the on-site Coulomb interaction through the $+U$ correction basically forces integer occupancy among 
only the correlated orbitals.

In addition, we find an increasing trend in the band gap with larger $U$ values or greater contributions of exact exchange to $V_{xc}$.
We find that $U=1.0$-2.0\,eV leads to very similar electronic structures obtained with HSE06 with 5-10\,\% exact exchange.
Our findings here suggest that a GGA$+U$ functional could be used as an alternative method to study electronic structures in lacunar spinels by reproducing the low-energy electronic structure obtained from a hybrid functional but at lower computational cost. The limitation is that lower lying orbitals that may be of interest are not corrected and therefore cannot exactly reproduce the results of the hybrid functional.
\revsecond{Based on our simulation results, we do not suggest using HSE06 functional for electronic structure simulations in the lacunar spinel family.}



\subsubsection{Optical conductivity}

\begin{figure}
  \centering
 \includegraphics[width=0.95\linewidth]{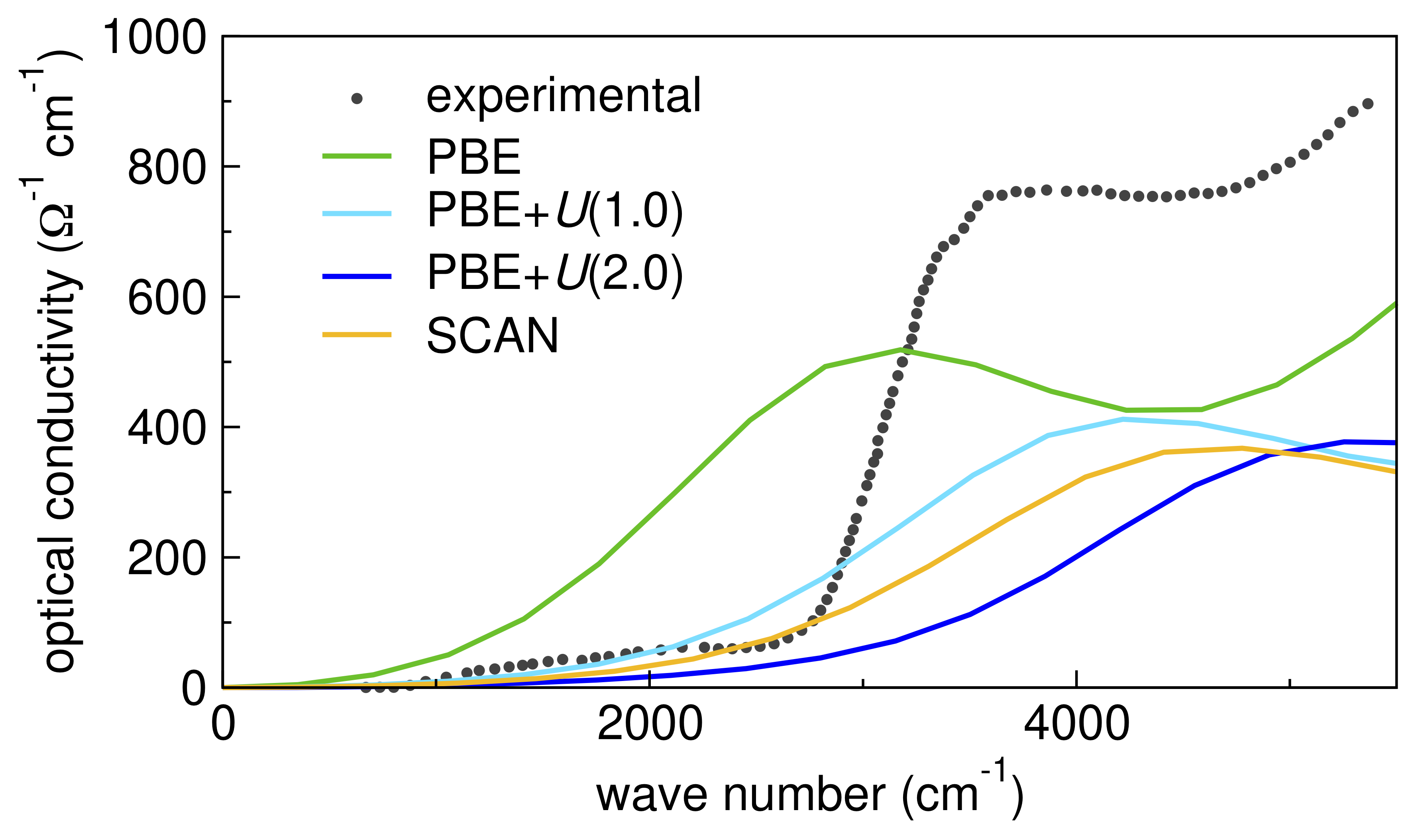}
  \caption{DFT calculated optical conductivity of \gvs{} in the rhombohedral $R3m$ structure compared with the experimental values obtained from Ref. \onlinecite{reschke2017optical}.}
  \label{fig:fig12}
\end{figure}

We compute the optical conductivity of the \rw{ferrimagnetic}  rhombohedral phase \gvs{}  
and compare our DFT results with the  experimental data \cite{reschke2017optical} in
\autoref{fig:fig12}.
The experimental data  shows the first optical transition occurs at $\approx$2,700\,cm$^{-1}$ (black symbols), corresponding to an approximate 0.33\,eV optical band gap. The optical conductivity then plateaus at approximately 800 $\Omega^{-1}\,\mathrm{cm}^{-1}$ for higher frequencies.
Our DFT simulations are able to semi-qualitatively capture the plateau structure, but do not quantitatively reproduce the optical conductivity.
With increasing values of $U$, the plateau shifts to higher frequency and this behavior coincides with a larger optical gap as expected from the aforementioned 
band gap dependencies on the exchange-correlation functional.
SCAN functional performs similar to PBE with $U=1.0$-$2.0$\,eV.
PBE with $U=2.0$\,eV gives an optical gap closest to the experimental value.
We note that because DFT is a single-particle ground state theory, it may not be the optimal tool to study excited state properties,
such as optical conductivity.
More accurate simulations, for example, could be pursued by solving the Bethe-Salpeter equation using the GW quasiparticle energies\cite{liu2018relativistic}.

\subsubsection{Lattice dynamics}

\begin{figure}
  \centering
 \includegraphics[width=0.95\linewidth]{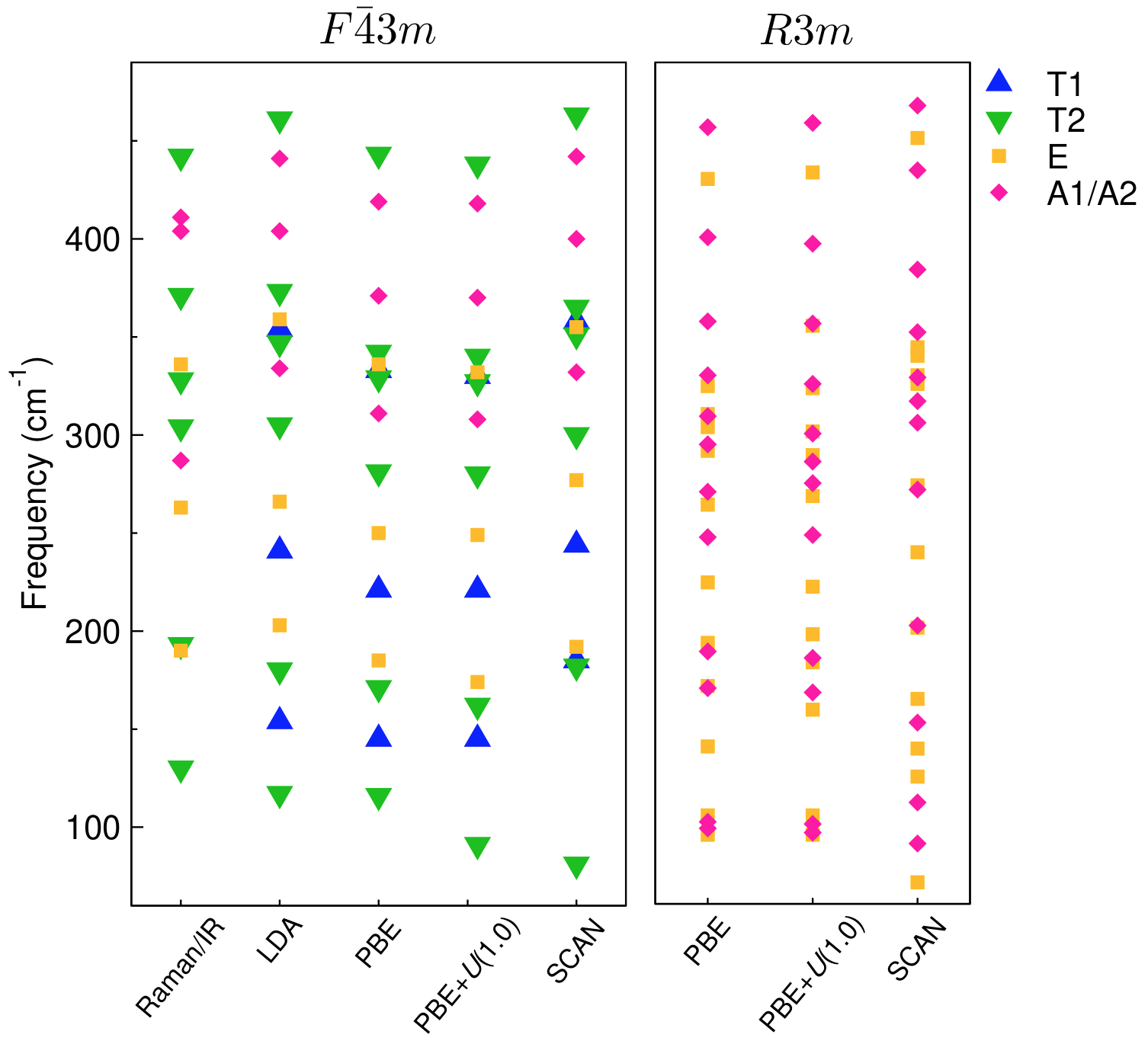}
  \caption{Phonon frequencies of \gvs{} in the cubic (left panel) and rhombohedral (right panel) phase with different DFT functionals. The experimental data from Ref.\ \onlinecite{hlinka2016lattice} is reproduced in the first column labeled `Raman/IR'.}
  \label{fig:fig13}
\end{figure}

Last, we investigate the exchange-correlation functional dependency on the phonon frequencies in \gvs{}.
We present our computed normal mode frequencies for both the cubic and rhombohedral phase with different functionals in 
\autoref{fig:fig13}: LDA, PBE, PBE$+U=1.0$\,eV, and SCAN. 
These calculated values are compared with the 
experimental Raman/IR frequencies at 80\,K reported in Ref.\ \onlinecite{hlinka2016lattice}, which 
appear in the first column of \autoref{fig:fig13}.
Note that the LDA results for the rhombohedral phase are not shown, because the $R3m$ structure is unstable at the 
LDA level.
We find that the cubic phase phonon frequencies generally agree well with experimental IR/Raman characterization data.
LDA and PBE perform reasonably well in reproducing the phonon frequencies.
However, with PBE$+U=1.0$\,eV, we find a significant decrease in the frequency of the lowest 
$T_2$ phonon mode.
The same behavior is also obtained with SCAN functional.  
Interestingly, this $T_2$ 
mode mainly corresponds to the distortion of the transition-metal cluster along the 
symmetry-breaking pathway.
This could be an evidence of electron-phonon coupling induced structure instability\cite{rastogi1984electron}.
Although no Raman/IR data is available for the rhombohedral phase, we still see the same phonon mode softening with functional choice upon going from PBE to SCAN.
The major difference between different functionals is at the low-frequency domain, 
where the vibrational modes are mainly related to the transition-metal (V$_4$) clusters.
Our finding here suggest that lattice dynamics in the lacunar spinels also have non-negligible functional dependency.
More experimental (temperature-dependent) data, however, is required to ascertain the functional that best reproduces the lattice dynamical properties.

\section{conclusions}

\rw{
In conclusion, LDA underperforms the other functionals and we recommended to use it only with on-site Coulomb interactions added.
The GGA functionals (PBE and PBEsol) perform reasonably well, and the results can be quantitatively improved with on-site Coulomb interactions explicitly added.
The meta-GGA functional SCAN is another alternative choice that works well and does not require extra parameterization.
Last, the hybrid functional HSE06 predicts accurate lattice structures, \revsecond{but} leads to a large electronic band gap
in the low-temperature rhombohedral phase. \revsecond{Owing to its high computational cost as well as large deviation in electronic structure predictions, we do not recommend using this hybrid functional for the lacunar spinel family.}}

\rw{
All exchange-correlation functionals predict reasonable lattice constants in both the cubic and rhombohedral polymorphs of the chalcogenide lacunar spinels.
For electronic structure simulations, 
the cubic phase is always metallic from band theory 
and exhibits a narrow transition-metal-derived bandwidth at the Fermi level.
Spin-orbit interactions are 
necessary to predict a semiconducting state in the cubic phase, 
but not in the rhombohedral phase, at the DFT level. 
At the LDA and GGA level, 
on-site Coulomb interactions of 2\,eV to 3\,eV are recommended to obtain quantitatively improved results.
We also found that the PBE functional without on-site Coulomb interactions could predict  stable semiconducting states for the rhombohedral phase.
Our results obtained with SCAN are similar to PBE$+U(2.0)$ and thus can be safely used in simulations.
%
We also find a highly spin-polarized DFT ground state in \gvs, which differs from available experimental data, motivating additional investigations of the magnetic order.
We found that the single-particle DFT simulations of the optical conductivity do not give a quantitatively satisfying description of \gvs; 
more sophisticated methods such as with the $GW$ method may be necessary to treat the excited state properties in the lacunar spinels.
The LDA and PBE functional, however, perform well in predicting cubic phase phonon frequencies in \gvs.
}


\begin{acknowledgments}
This work in supported by the National Science Foundation (NSF) under award number DMR-1729303. 
\textit{Ab initio} DFT simulations are performed on Extreme Science and Engineering Discovery Environment (XSEDE), 
which is supported by NSF grant number ACI-1548562
and the DoD-HPCMP (Copper cluster). 
The authors thank E.\ Schueller, J.\ Zuo, R.\ Seshadri, and S.\ Wilson for useful discussions.

\end{acknowledgments}

\bibliography{reference}

\end{document}